\documentclass[11pt]{article}
\usepackage{epstopdf}                                        
\usepackage[a4paper,hmarginratio=1:1,vmarginratio=2:3,totalwidth=15.3cm,totalheight=22.54cm,
]{geometry}
\usepackage{bm,epstopdf,epsfig,amsmath,amssymb,amsfonts,colordvi,latexsym,
comment,cancel,verbatim,slashed}
\usepackage{epsfig,bbm}
\usepackage{graphicx,graphics}
\usepackage[font=md,captionskip=8pt]{subfig}
\usepackage[usenames,dvipsnames]{color}
\usepackage[noadjust]{cite}
\usepackage{xcolor} 
\usepackage[utf8]{inputenc}
\usepackage{setspace}
\newcommand\inlineeqno{\stepcounter{equation}\ (\theequation)}
\newcommand{\sg}{\sqrt{g}}    
\newcommand{\sqg}{\sqrt{g}}
\newcommand{\sqgh}{\sqrt{\hat g}}
\newcommand{\gh}{\hat g} 
\newcommand{\w}{v}

\newcommand{\q}{\alpha}
\newcommand{\h}{\sigma}

\newcommand{\tGamma}{\tilde\Gamma}
\allowdisplaybreaks

\newcommand{\cO}{{\cal O}}

\newcommand{\cV}{{\mathcal V}}

\newcommand{\ra}{\rightarrow}

\newcommand{\be}{\begin{equation}}
\newcommand{\ee}{\end{equation}}
\newcommand{\bea}{\begin{eqnarray}}
\newcommand{\eea}{\end{eqnarray}}

\newcommand{\baa}{\begin{array}}
\newcommand{\eaa}{\end{array}}

\long\def\symbolfootnote[#1]#2{\begingroup
\def\thefootnote{\fnsymbol{footnote}}\footnote[#1]{#2}\endgroup}
\begin{document} 
\begin{flushright}
\end{flushright}
\bigskip\medskip
\thispagestyle{empty}
\vspace{2cm}
\begin{center}

 {\Large \bf  Palatini quadratic gravity: spontaneous breaking 

\bigskip
of gauged scale symmetry and inflation}

\vspace{1.cm}

 {\bf D. M. Ghilencea} \symbolfootnote[1]{E-mail: dumitru.ghilencea@cern.ch}

\bigskip
{\small Department of Theoretical Physics, National Institute of Physics
 
and Nuclear Engineering, Bucharest\, 077125, Romania}
\end{center}

\bigskip
\begin{abstract}
\begin{spacing}{1.07}
\noindent
We study quadratic gravity $R^2+R_{[\mu\nu]}^2$ 
in  the Palatini formalism where the  connection and the metric are independent.
This  action has a {\it gauged} scale symmetry (also known as Weyl gauge symmetry) of Weyl 
gauge field $\w_\mu= (\tGamma_\mu-\Gamma_\mu)/2$, with $\tGamma_\mu$ ($\Gamma_\mu$)  the trace
of the Palatini (Levi-Civita) connection, respectively.
The underlying geometry is non-metric due to the $R_{[\mu\nu]}^2$ term acting
as a gauge kinetic term for $\w_\mu$.
We show that  this theory has an elegant spontaneous breaking of gauged scale symmetry
and mass generation in  the absence of matter, where the
necessary scalar field ($\phi$) is not added ad-hoc to this purpose
but is ``extracted'' from the $R^2$ term.
The gauge field becomes massive by absorbing the derivative term
$\partial_\mu\ln\phi$ of the Stueckelberg field (``dilaton''). 
In the broken phase one finds the Einstein-Proca action of $\w_\mu$
of mass proportional to the Planck scale $M\sim \langle\phi\rangle$, and
a positive cosmological constant.
Below this scale $\w_\mu$  decouples, the connection 
becomes Levi-Civita and metricity and Einstein gravity are recovered.
These results remain valid in the presence of non-minimally coupled scalar field (Higgs-like)
with  Palatini connection and the potential is computed.
In this case   the theory
gives successful inflation and a  specific prediction for the tensor-to-scalar ratio 
$0.007\!\leq\! r\!\leq 0.01$ for current spectral index  $n_s$ (at $95\%$CL) and
$N=60$ efolds. This value of $r$ is mildly larger than in inflation in  Weyl quadratic
gravity of similar symmetry,
due to  different  non-metricity. This establishes a connection between non-metricity 
and inflation predictions and enables us to test such theories by future CMB 
experiments.
\end{spacing}
\end{abstract}

\newpage

\section{Introduction}

At a fundamental level gravity may be regarded as a theory of connections. 
An example is  the ``Palatini approach''  to gravity due to  Einstein
\cite{E1,E2}, hereafter called EP approach \cite{c7,So1}. 
In this case the  ``Palatini connection'' ($\tGamma$) 
is apriori independent of the metric ($g_{\alpha\beta}$) and 
is  actually determined by its equations of motion, from the action considered.
For simple actions, $\tGamma$  plays  an auxiliary role only, with no dynamics. 
For example, for an Einstein action  in the EP approach the variation principle  
gives  that $\tGamma$ is actually equal to  the Levi-Civita connection ($\Gamma$).  
With this solution for $\tGamma$,  one  then  recovers Einstein 
gravity -  the  metric formulation and EP approach  are  equivalent.

However, this  equivalence is not true in general, for complicated actions,  
with matter present, etc \cite{cc1,Vollick2,c1,c2,c3,c4,c5,c6,c8,c9,Kozak,VV0,SO1,Vollick1,Annala,Percacci3}.
For example, for  quadratic gravity actions of the type  studied here
in the EP approach, the equations of motion of $\tGamma$ become complicated
 {\it  second-order}  differential
equations;  further,  some components of $\tGamma$ even  become dynamical in a sense
discussed shortly, etc. The question remains, however, if  such general actions  in the EP formalism
and {\it in the absence of matter} can recover {\it dynamically}  the Levi-Civita  
connection and Einstein  gravity. 
If true, this would be similar to the original Weyl quadratic  gravity theory
\cite{Weyl1,Weyl2,Weyl3,Scholz}  as we showed recently in \cite{Ghilen1,Ghilen2}.
The  main goal of this paper is to answer this question.

To address this question we study a gravity action in the EP approach
with {\it gauged} scale symmetry also called Weyl gauge symmetry,
see \cite{Ghilen1,Ghilen2} for an example\footnote{See e.g.\cite{Dirac,
Ross,Scholz,Smolin,Cheng,JW,Quiros,Moffat1,Oh,Moffat2,ghilen, Heisenberg,Jimenez,Jimenez2,
Hill,P,Scholz2,Dengiz,Tann,Fulton,Weyl1,Weyl2,Weyl3,Ghilen1,Ghilen2,Winflation}
for models with gauged scale symmetry and 
\cite{Turok,tH0,TH,tH1,sS1,Armillis,Kubo,
higgsdilaton,Monin,D1,Lalak2,D2,K1,R1,FRH1,FRH2,CW-BL,lebedev,Lalak,Strumia,SO2,SO3}
 for  conformal or global scale symmetry.}. This symmetry, first present in Weyl gravity \cite{Weyl1,Weyl2,Weyl3}
is  important for  mass  generation, hence our interest.
This symmetry demands us to  consider   quadratic gravity actions, with 
no dimensionful  parameters.
For such action we  shall:  
{\bf 1)}  explain the spontaneous breaking of this symmetry and the 
 emergence  of Levi-Civita connection,  Einstein gravity 
 and Planck scale in the broken phase,
even in the absence of matter. This  answers the above question;
{\bf 2)} study the relation of this action to  Weyl theory 
\cite{Weyl1,Weyl2,Weyl3} of similar symmetry; 
{\bf 3)}  study its inflation predictions.

In Section~\ref{2} we first review the $R(\tGamma,g)^2$ gravity in the EP approach, where
$R(\tGamma,g)$ denotes the scalar curvature in this formalism.
This action is local scale invariant. 
The connection is shown to be conformally related to the Levi-Civita 
connection. When ``fixing the gauge'' of this symmetry, the ``auxiliary'' scalar field 
$\phi$ introduced to ``linearise'' the $R^2$ term decouples.
As a result, one finds that   $\tGamma=\Gamma$ and Einstein action is obtained.

In Sections~\ref{3} and \ref{4} we study the
quadratic action   $ R(\tGamma,g)^2+R_{[\mu\nu]}(\tGamma)^2$ in the EP approach,
 hereafter called ``EP quadratic gravity''.  Here we used the notation
 $R_{[\mu\nu]}\!\equiv \! (R_{\mu\nu}-R_{\nu\mu})/2$.
In this action the trace $\tGamma_\mu$ of the Palatini connection (assumed symmetric) 
is  {\it dynamical} in the sense that 
$R_{[\mu\nu]}(\tGamma)^2$ is a  gauge kinetic  term for $\tGamma_\mu$ or,  
more exactly, for the vector field\footnote{Unlike $\tGamma_\mu$ and $\Gamma_\mu$, 
 $\w_\mu\propto\tGamma_\mu-\Gamma_\mu$ is indeed a vector 
(see Appendix). We  assume  $\tGamma_{\mu\nu}^\alpha=\tGamma_{\nu\mu}^\alpha$ (no torsion).}
 $\w_\mu\!\sim\! \tGamma_\mu-\Gamma_\mu$, \,
($\tGamma_\mu\!\equiv\! \tGamma^\alpha_{\mu\alpha}$,\,
$\Gamma_\mu\!\equiv\! \Gamma^\alpha_{\mu\alpha}$).
With  $\tGamma$   independent of $g_{\mu\nu}$, one notices that
 the local scale symmetry of this   action is actually
 a  {\it gauged} scale symmetry, of gauge field  $\w_\mu$.

A consequence of the gauged scale symmetry is that 
 EP quadratic gravity is  non-metric\footnote{ 
Non-metricity means  that under parallel transport  along a curve  a vector  changes its
 norm i.e. is path dependent; therefore, in realistic theories with matter present (as here)
 it must be suppressed  by a large scale (e.g.Planck) to avoid  atomic spectral 
lines changes that it would otherwise induce
 \cite{Weyl1}. In the absence of matter  non-metricity can be traded for
torsion in related $R^2$ theory \cite{JJ2}. } i.e. 
 $\tilde\nabla_\mu g_{\alpha\beta}\!\not=\!0$. This is 
due to a  dynamical $\w_\mu\sim \tGamma_\mu$
and $\w_\mu$ is the  non-metricity field, also called Weyl gauge field. 
Further, we find that the  equations  of motion for $\tGamma$ are 
 {\it second-order} differential equations.
In this case  the usual EP approach in $f(R)$ theories  to solve algebraically for
 $\tGamma$  \cite{So1} does not work, due to local scale symmetry and non-metricity.
Nevertheless, we  compute $\tGamma$ and find
 that  EP quadratic gravity  with $\tGamma$ {\it onshell}
is equivalent to a ghost-free second-order gauged scale invariant theory with an {\it additional}
dynamical field (``dilaton''). (Expressed in terms of this field 
 the differential equations of $\tGamma$ simplify considerably and this is how they are 
solved).

The main  result of this work   (Section~\ref{3}) is that
the gauged scale invariance of the above action is broken spontaneously
by a new mechanism  \cite{Ghilen1,Ghilen2}  valid even in the absence of matter;
in this,  the necessary scalar field ($\phi$)  is {\it not} added ad-hoc to this purpose
(as usually done), but is ``extracted'' from the $R^2$ term in the action;
$\phi$ is thus of geometric origin.
After  a  Stueckelberg mechanism \cite{S1,P1,P2}  the
gauge field $\w_\mu$ becomes massive, of mass $m_\w$ near the Planck scale $M\sim \langle\phi\rangle$,
by ``absorbing'' the  derivative  $\partial_\mu\ln\phi$ of the Stueckelberg field
(also referred to as ``dilaton'').
Near the  Planck scale we obtain the Einstein-Proca action of $\w_\mu$.
Further, below the scale  $m_\w\!\propto\!  M$, the field $\w_\mu$ decouples and we recover  metricity,
 Levi-Civita connection  and  Einstein gravity;
the Planck scale $M$ is then an {\it emergent} scale  where  this symmetry is broken.
These results  remain true if the theory also has matter  fields  (higgs,  etc) 
non-minimally coupled with {\it Palatini connection}, while respecting  gauged scale invariance
(Section~\ref{4}). Briefly,  the  EP quadratic gravity
is a gauged scale invariant theory broken \`a la Stueckelberg, even in the absence of matter,
to an Einstein-Proca action with a positive cosmological constant and
a potential for the scalar fields - if present.
This  answers  the main goal  of the paper.

Another theory where the connection is not determined by the metric itself is 
the original  Weyl quadratic gravity of gauged scale invariance
 \cite{Weyl1,Weyl2,Weyl3} (also \cite{Scholz}).
 With hindsight, it is then not too surprising that the above results are similar to 
those in \cite{Ghilen1,Ghilen2,Winflation}  for  Weyl theory.
This  theory came under early criticism from Einstein \cite{Weyl1} for its 
non-metricity implying e.g. changes of the atomic spectral lines, in contrast to experiment; 
however, if the Weyl ``photon'' ($\w_\mu$) of non-metricity is actually massive (mass $\sim\! M$)
by  the  same Stueckelberg mechanism,  metricity and Einstein gravity are  
recovered below its decoupling scale ($\sim$ Planck scale).
Non-metricity effects are then strongly suppressed by a large $M$ (their current lower bound 
seems  low \cite{Latorre,Lobo}). Hence, the long-held criticisms that have implicitly
 assumed $\w_\mu$ be massless
are actually avoided and Weyl gravity is then viable  \cite{Ghilen1,Ghilen2,Winflation}. 
As outlined, in this work we obtain similar results in  EP quadratic gravity,
up to different  non-metricity effects.

We also  study inflation in  EP quadratic gravity (Section~\ref{5}).
We consider this theory with an extra scalar field (Higgs-like)
with perturbative non-minimal coupling and Palatini connection, that plays the role of the inflaton.
We  compute the potential after the gauged scale symmetry breaking.
With the Planck scale  a simple phase transition  scale in our theory, field values 
above $M$ are natural. Interestingly, the inflaton potential  is similar to that  in
 Weyl quadratic gravity  \cite{Winflation}, 
up to couplings and field redefinitions  (due to a different non-metricity of the theory).
Inflation in EP quadratic gravity has a  specific prediction for the
 tensor-to-scalar ratio $0.007\leq r \leq 0.010$  for the current spectral index $n_s$ at $95\%$\,CL. 
This range of $r$ is distinct from that predicted by inflation in  Weyl gravity \cite{Winflation,Ross}
and will soon be reached by  CMB experiments \cite{CMB1,CMB2,CMB3}. 
The Conclusions are presented in Section~\ref{6}
followed by an Appendix.

\section{Palatini $R^2$ gravity}\label{2}

For later reference we first review  $R^2$ gravity  in the EP formalism 
\cite{Englert,Hi}.
As discussed below, the  action is local scale invariant (unlike its Riemannian counterpart):
\bea\label{e01}
L_1=\sqg\, \, \frac{\xi_0}{4!}\,
 R(\tilde \Gamma,g)^2, \qquad  \xi_0>0,
\eea
where
\bea\label{e02}
R(\tGamma,g)=g^{\mu\nu}\,R_{\mu\nu}(\tGamma),\qquad
R_{\mu\nu}(\tGamma)=\partial_\lambda\tilde\Gamma^\lambda_{\mu\nu}-
\partial_\mu \tGamma^\lambda_{\lambda\nu}
+\tGamma_{\rho\lambda}^\lambda \tGamma^\rho_{\mu\nu}-\tGamma^\lambda_{\rho\mu}
\tGamma^\rho_{\nu\lambda}
\eea
$R_{\mu\nu}(\tGamma)$ is the metric-independent Ricci tensor in the EP formalism.
Our conventions are as in \cite{book} with  metric
 (+,-,-,-), \,\, $g\equiv \vert \det g_{\mu\nu}\vert$ and  we
assume there is no torsion i.e. $\tGamma_{\mu\nu}^\rho=\tGamma_{\nu\mu}^\rho$.

There is an  equivalent ``linearised''  version of $L_1$, found  by using an auxiliary field $\phi$
\medskip
\bea\label{e03}
L_1=\sqg\, \, \frac{\xi_0}{4!}\,
\Big\{-2 \,\phi^2 R(\tilde \Gamma,g)-\phi^4\Big\}.
\eea

\medskip\noindent
Indeed, (\ref{e01}) is recovered if we use  in (\ref{e03}) the solution  $\phi^2=-R(\tGamma,g)$ of the
equation of motion of the scalar field $\phi$. With the  connection $\tilde\Gamma$
independent of the metric,   (\ref{e03}) and (\ref{e01}) have local scale symmetry
  i.e. are  invariant under a Weyl transformation  $\Omega=\Omega(x)$ with\footnote{
From $\phi^2=-R(\tGamma,g)$,  $\phi^2$ transforms under metric rescaling 
like $R(\tGamma,g)$, as expected for a scalar field.}
\medskip
\bea
\label{cr}
\hat g_{\mu\nu}=\Omega^2 g_{\mu\nu},\quad
\sqgh=\Omega^4\sqg,\quad
\quad \hat\phi=\frac{1}{\Omega}\phi,
\quad 
\hat R(\tGamma,\gh)=\frac{1}{\Omega^2} R(\tGamma,g).
\eea

\medskip \noindent
Unlike in the  metric case, $R_{\mu\nu}(\tGamma)$ is invariant under (\ref{cr}) while
$R(\tGamma,g)$ transforms covariantly, hence (\ref{e01}), (\ref{e03}) are invariant.
$L_1$  has a shift symmetry:  $\ln\phi\!\ra\!\ln\phi-\ln\Omega$.
In global cases  $\ln\phi$ is the  dilaton field generating a mass scale from
its vev (assumed to be non-zero);
here,  $\ln\phi$ is similar to a would-be Goldstone, as seen
 if we  ``gauge''  symmetry (\ref{cr})  (see later,  eq.(\ref{s2})).

Let us solve the equation of motion for $\tGamma$, then find the action for $\tGamma$ 
{\it onshell} \footnote{
Obviously, with $\Omega^2=\xi_0 \phi^2/(6 M^2)$ (with $M$ the Planck scale), one
can set  $\phi$ to a constant (fix the ``gauge'' of local scale symmetry).
$L_1$ becomes  $L_1=\sqrt{-g}\, \big\{ - (1/2) M^2 \,R(\tGamma,g)-3/(2 \xi_0) M^4\big\}.$
This is  the  Palatini formulation of Einstein action;  via eqs motion 
then $\tilde\nabla_\mu g_{\alpha\beta}=0$ where $\tilde\nabla_\mu$ is computed with
$\tGamma$. Hence $\tGamma$ is  a Levi-Civita  connection.
 However, this approach obscures the role of local scale symmetry, relevant later.}. 
The change of $R_{\mu\nu}(\tGamma)$ under a variation of the connection is 
$\delta R_{\mu\nu}(\tGamma)=\tilde\nabla_\lambda(\delta\tGamma_{\mu\nu}^\lambda) -
\tilde\nabla_\nu(\delta\tGamma_{\mu\lambda}^\lambda)$,\,  
where the operator $\tilde\nabla$ is defined with connection $\tGamma$.
Then from (\ref{e03}) the equation of motion of $\tGamma_{\mu\nu}^\lambda$ gives
\bea\label{eq01}
\tilde \nabla_\lambda (\sqg\, g^{\mu\nu}\phi^2 ) 
- \frac12\,\Big[
 \tilde\nabla_\rho (\sqg \,g^{\rho\mu} \phi^2)\, \delta_\lambda^\nu+(\mu\leftrightarrow \nu)\Big]
=0.
\eea

\medskip\noindent
Setting $\nu=\lambda$ and then summing over, then\footnote{If we use 
$\phi^2=-R(\tGamma,g)$ in (\ref{eq01}), (\ref{s0}),
 one recovers the equation of motion of $\tGamma$ found
 directly from (\ref{e01}) \cite{Englert}.}
\bea\label{s0}
\tilde\nabla_\rho(\sqg \,g^{\rho\mu}\phi^2)=0.
\eea

\medskip\noindent
To simplify notation, introduce an auxiliary  dimensionful  ``metric'' 
$h_{\mu\nu}\equiv \phi^2 g_{\mu\nu}$, then
\bea\label{rel0}
\tilde\nabla_\lambda(\sqrt{h}\, h^{\mu\nu})=0.
\eea
%
This  means that in terms of $h_{\mu\nu}$, the connection is  
Levi-Civita\footnote{One shows  $\tilde\nabla h_{\mu\nu}=0$ by using
$\label{rel1}
\quad
 \tilde\nabla_\lambda h^{\mu\nu}
= -h^{\mu\sigma} h^{\nu\rho}\tilde \nabla_\lambda h_{\sigma\rho},
\,\, \textrm{and}\,\,\,
\tilde\nabla_\lambda\sqrt{h} =(1/2) \,\sqrt{h}\, h^{\alpha\beta}\tilde \nabla_\lambda h_{\alpha\beta}.
\hfill{\inlineeqno}$}
\bea\label{oonn}
\tGamma^\alpha_{\mu\nu}(h)=(1/2)\, h^{\alpha\lambda} ( \partial_\mu h_{\lambda\nu}
+\partial_\nu  h_{\lambda\mu}-
\partial_\lambda h_{\mu\nu}),
\eea
\vspace{-0.8cm}

\noindent
or, in terms of $g_{\mu\nu}$
\bea\label{on}
\tGamma_{\mu\nu}^\alpha=\Gamma_{\mu\nu}^\alpha(g)
+
(1/2)\,
\big(\delta_\nu^\alpha \,u_\mu+\delta_\mu^\alpha u_\nu
-g^{\alpha\lambda} g_{\mu\nu} u_\lambda\big),\quad u_\mu\equiv\partial_\mu \ln\phi^2,
\eea
%
with  Levi-Civita 
$\Gamma^\alpha_{\mu\nu}(g)=(1/2) g^{\alpha\lambda} (\partial_\mu g_{\lambda\nu}+\partial_\nu g_{\lambda\mu}
-\partial_\lambda g_{\mu\nu})$.
Next, if we  use the equation of motion of $\phi$ of solution $\phi^2\!=\!-R(\tGamma,g)$,
 eq.(\ref{on}) for $\tGamma$ (also (\ref{eq01}), (\ref{s0}))  becomes
a {\it second-order} differential equation
since $\partial\phi^2\sim\partial R\sim\partial^2\tGamma$,  and it is difficult to solve
(and since solution $\tGamma$ of (\ref{on}) involves  $\partial g_{\mu\nu}$  from $\Gamma(g)$
then for $\tGamma$ onshell action  (\ref{e01}) is a {\it four-derivative}
theory in $g_{\mu\nu}$).  An  easy way out is to keep $\phi$
 an independent variable hereafter (no use of its eq of motion), then 
eqs.(\ref{eq01}), (\ref{s0})  have solution $\tGamma$ given by the rhs of  (\ref{on}). 
For this solution, then
\bea\label{rel}
R(\tGamma,g)=R(g)- 3 \nabla_\mu u^\mu-\frac{3}{2} \, g^{\mu\nu} u_\mu \, u_\nu,
\eea
with the  Ricci  scalar $R(g)$ for $g_{\mu\nu}$ while $\nabla$ is defined  with the 
Levi-Civita connection ($\Gamma$). Using (\ref{rel}) in (\ref{e03}) of the same metric, we find
for $\tGamma$ onshell\footnote{From (\ref{rt}) the equation of motion for $g^{\mu\nu}$
and its trace give that on the  ground state $\phi^2=-R$ and 
$R (R_{\mu\nu}- 1/4 g_{\mu\nu} R)=0$ which is similar to 
that found directly from  equivalent  action (\ref{e01}), see also footnote 6.}
\bea\label{rt}
L_1=\sqg\,  \Big\{\frac{\xi_0}{2}\,\Big[- \frac{1}{6}\, \phi^2 R(g) -  (\partial_\mu\phi)^2
-\frac{1}{12} \,\phi^4\Big]\Big\}.
\eea
%
$L_1$  is a second order theory     with an additional dynamical variable
 demanded by symmetry (\ref{cr}) and is  equivalent to action (\ref{e01}) which  for $\tGamma$ onshell
 is a four-derivative theory, as noticed.

Lagrangian (\ref{rt})  has  local scale symmetry
so one may like to ``fix the gauge''.  We  choose the Einstein or unitarity gauge
reached by  a $\phi$-{\it dependent} 
transformation   $\Omega^2\!=\!\phi^2\!/\langle\phi\rangle^2$
 that is gauge-fixing  $\phi$  to a {\it constant} (vev);
in this gauge  $M^2\!\!=\!\xi_0\langle\phi\rangle^2/6$ is the Planck mass.
From~(\ref{rt})
\bea\label{ss}
L_1=\sqgh \, \Big\{ \frac{-1}{2} M^2 \hat R(\hat g) - \frac{3}{2 \xi_0} M^4\Big\}. \quad
\eea
Hence Einstein action (\ref{ss})  is recovered  {\it as a gauge fixed form} of  (\ref{rt});
 symmetry (\ref{cr}) is now spontaneously broken and  $\phi$ decouples\footnote{
There is no gauge field here to ``eat'' would-be Goldstone $\ln\phi$
but in Section~\ref{3},  $\w_\mu$ of (\ref{s2})  will ``eat'' $\ln\phi$.}
 \cite{Nakayama};
this may be expected since the local scale symmetry current
of (\ref{rt}) is vanishing \cite{J1,J2,tHooft} 
(this will change in Section~\ref{3.3}). With $\phi$ ``gauge fixed'' to a constant,
eqs.(\ref{rel0}),(\ref{on})  give $h_{\mu\nu}\!\propto\! g_{\mu\nu}$ and $\tGamma\!=\!\Gamma$ 
so the theory is metric\footnote{This result is also valid
 for Palatini $f(R)$ action instead of  (\ref{e01}); we do not 
consider it here since  it violates  Weyl scale  symmetry,
but (unlike here) the trace of the equation of motion of $g^{\mu\nu}$ 
is non-trivial, giving $f'(R)\!=$constant, which fixes  $R$,
then $h_{\mu\nu}\!\propto\! g_{\mu\nu}$,  $\tGamma=\Gamma$ so
 metricity/Einstein action is recovered\cite{So1,c7}. }$^,$\footnote{
This situation is different from the Riemannian $R(g)^2$  gravity \cite{K}
 (eq.2.11), also \cite{LAG}, where in the Einstein frame a kinetic term for 
$\phi$ remains  present  in (\ref{ss}) (being absent in eq.(\ref{rt})).}.

\section{Palatini  quadratic gravity with gauged scale symmetry}\label{3}

\subsection{The Lagrangian and its expression for onshell $\tGamma$}

Consider now the following  EP quadratic gravity, with $\alpha$=constant and 
$R_{[\mu\nu]}\!\equiv\! (R_{\mu\nu}-R_{\nu\mu})/2$
\bea\label{qq1}
\qquad
L_2=\sqg\, \Big\{\,\frac{\xi_0}{4!}\,R(\tGamma,g)^2- \frac{1}{4 \q^2}
 R_{[\mu\nu]}(\tGamma)\, R^{\mu\nu}(\tGamma)
\Big\}.
\eea

\medskip\noindent
 With $R_{\mu\nu}(\tGamma)$ from eq.(\ref{e02}) 
and $\tilde\Gamma^\alpha_{\mu\nu}$   symmetric in $(\mu,\nu)$,\,
$L_2$ has  a more intuitive form
\medskip
\bea\label{e1}
L_2=\sqg\, \Big\{\, \frac{\xi_0}{4!}\,
 R(\tilde \Gamma,g)^2 -\frac{1}{4 \q^2}\, 
F_{\mu\nu}(\tilde \Gamma) F^{\mu\nu}(\tilde\Gamma)\Big\}.
\eea

\medskip\noindent
This  is a natural extension of $L_1$ of eq.(\ref{e01}),  with 
the second term above indicating  we now  have a  dynamical trace ($\tGamma_\mu$) 
of the Palatini  connection, as seen from the notation below:
\medskip
\bea\label{fmunu}
F_{\mu\nu}(\tGamma)=
 \tilde\nabla_\mu \w_\nu -\tilde\nabla_\nu \w_\mu;
\qquad
\w_\mu= (1/2)\big(\tGamma_\mu-\Gamma_\mu(g)\big),
\eea

\medskip\noindent
with  $\tGamma_\mu\equiv \tGamma_{\mu\lambda}^\lambda$ and
$\Gamma_\mu\equiv \Gamma_{\mu\lambda}^\lambda$.
Since  $\tilde\Gamma^\alpha_{\mu\nu}=\tGamma_{\nu\mu}^\alpha$  and 
$\tilde\nabla_\mu \w_\nu=\partial_\mu v_\nu - \tGamma_{\mu\nu}^\alpha v_\alpha$,
 then we have  $F_{\mu\nu}=\!\partial_\mu \w_\nu-\partial_\nu \w_\mu
=(\partial_\mu\tGamma_\nu-\partial_\nu\tGamma_\mu)/2=-R_{[\mu\nu]}$, and
 eqs.(\ref{qq1}), (\ref{e1}) are equivalent.
While $\Gamma_\mu(g)$  does not contribute to $F_{\mu\nu}(\tGamma)^2$, it is needed to
ensure that  $\w_\mu$  is a vector under  coordinate transformation
(which is not true for  $\tGamma_\mu$ or $\Gamma_\mu$, see Appendix).
$\w_\mu$ is the  Weyl field\footnote{Definition (\ref{fmunu})
of  gauge field $\w_\mu$ is general, it also applies to Weyl gravity of similar symmetry
(Appendix).} and measures the trace of the deviation 
of the Palatini connection   $\tGamma$ from  Levi-Civita   connection $\Gamma(g)$. 
$L_2$ is  quadratic in $R$ but for $\tGamma$ offshell resembles  a second order theory.

As in previous section,  write $L_2$ in an equivalent ``linearised''  form useful later on
\medskip\noindent
\bea\label{sss}
L_2=\sqg\, \Big\{- \frac{\xi_0}{12}\, \phi^2 \,R(\tilde\Gamma,g)
- \frac{1}{4 \q^2}\,  F_{\mu\nu}(\tGamma)^2
-\frac{\xi_0}{4!}\,\phi^4
\Big\}.
\eea

\medskip\noindent
The equation of motion for $\phi$ has  solution
$\phi^2\!=\!-R(\tGamma,g)$ which replaced in $L_2$ recovers (\ref{e1}).

Since $\tGamma$ does not transform under (\ref{cr}) and with  
$\Gamma_\mu(g)\!=\!\partial_\mu\ln\sqg$ that follows from the definition of 
Levi-Civita connection, then $L_2$   is invariant under (\ref{cr}) extended by
\bea\label{s2}
\hat \w_\lambda=\w_\lambda-\partial_\mu \ln \Omega^2.
\eea

\medskip\noindent
The invariance of $L_2$  under transformations (\ref{cr}), (\ref{s2}), 
is referred to  as {\it gauged} scale invariance or Weyl gauge symmetry, 
with a (dilatation) group  isomorphic to ${\bf R^+}$, as in Weyl  gravity.

Let us then compute the connection $\tGamma_{\mu\nu}^\lambda$ from its equation of motion
which is
\medskip\bea\label{eom1}
\tilde\nabla_\lambda(\sqg\,g^{\mu\nu}\phi^2)
-\Big\{
\frac12\,\delta_\lambda^\nu\Big[
 \tilde\nabla_\rho (\sqg\, g^{\mu\rho} \phi^2)
{-}\frac{6 \sqg}{\q^2 \xi_0} \nabla_\rho F^{\rho\mu}\Big]
+(\mu\leftrightarrow\nu)\Big\}=0.
\eea

\medskip\noindent
Here $\tilde\nabla_\mu$ and $\nabla_\mu$ are evaluated with the Palatini ($\tGamma$)  and 
Levi-Civita ($\Gamma$) connections, respectively. Setting $\lambda=\nu$ and 
summing over gives  (compare against eq.(\ref{s0}))
\bea\label{t0}
\tilde\nabla_\rho(\sqg\,g^{\mu\rho}\,\phi^2)
=\frac{10}{ \q^2} \frac{1}{\xi_0} \sqg\, \nabla_\rho F^{\rho\mu},
\eea

\smallskip\noindent
which is an equation of motion for the trace $\tGamma_\mu\sim \w_\mu$. 
Replacing (\ref{t0}) back in (\ref{eom1})  leads to
\medskip
\bea\label{eq23}
\tilde\nabla_\lambda(\sqg\,g^{\mu\nu}\phi^2)
-\frac15\,
\Big\{
\,\delta_\lambda^\nu\,
\tilde\nabla_\rho (\sqg\, g^{\rho\mu}\phi^2)
+(\mu\leftrightarrow\nu)\Big\}=0.
\eea

\medskip\noindent
Therefore, the set of eqs.(\ref{eom1}) is equivalent to the combined set of 
eqs.(\ref{eq23}) and (\ref{t0})\footnote{
Unlike in (\ref{eom1}), setting $\lambda=\nu$ in (\ref{eq23}) brings no information  - this
was ``moved''  into  4 eqs in (\ref{t0}).}.

Let us find $\tGamma_{\mu\nu}^\lambda$ from (\ref{eq23}).
Note that if one  used  the equation of motion of $\phi$ of solution
$\phi^2= -R(\tGamma,g)$,  then (\ref{eq23}) would be  a {\it second-order} differential equation 
for $\tGamma^\alpha_{\mu\nu}$, since  $\tilde\nabla_\lambda\phi^2\sim
\partial\phi^2\sim\! \partial R(\tilde\Gamma,g)\sim \partial^2\tGamma$, 
with  further complications. It is however easier to simply
regard $\phi$ hereafter as an independent variable\footnote{The consequence of doing so
 is that $\phi$   acquires a kinetic term and becomes dynamical (see also Section~\ref{2}).}
(i.e. no use of its equation of motion)
 in terms of which one 
then easily computes $\tGamma$ algebraically,
 as we do below.
 To find a solution to (\ref{eq23}) we first introduce, based on an approach of \cite{Vollick2}:
\medskip
\bea\label{tt1}
\tilde\nabla_\lambda(\sqrt{g}\,g^{\mu\nu}\phi^2)
=(-2)\sqrt{g} \,\phi^2(\delta_\lambda^\mu \, V^\nu +\delta_\lambda^\nu\,V^\mu),
\eea

\medskip\noindent
where $V_\mu$ is   some arbitrary vector field (to be determined later). $V_\mu$
 is introduced since, due to  underlying symmetry, eq.(\ref{eq23}) with $\lambda=\nu$ summed over
  is automatically respected for fixed $\mu$ ($=0,1,2,3$); this is
leaving four undetermined components, accounted for by $V_\mu$.
Further, if in eq.(\ref{eq23}) one replaces $\tilde\nabla (..)$ terms by the rhs of (\ref{tt1}) 
one easily shows  that (\ref{eq23}) is indeed verified.
Hence, instead of finding $\tGamma$ from (\ref{eq23}), it is sufficient to
compute  $\tGamma$ from (\ref{tt1})\footnote{
One cannot solve  algebraically (\ref{tt1})  as done in Palatini $f(R)$ 
theories \cite{c7,So1} due to non-vanishing rhs (dynamical $\tGamma_\mu$) and to 
the conformal symmetry of $L_2$, absent in  $f(R)$ theories, 
see discussion in  \cite{VV}, p.5-6.}, which is easier.
To this end, multiply (\ref{tt1}) by $g_{\mu\nu}$ and use 
$g_{\mu\nu}\tilde\nabla_\lambda g^{\mu\nu}=-2 \tilde\nabla_\lambda \ln\sqrt{g}$, to find
that  
\bea\label{t2}
V_\lambda=- (1/2)\,\tilde\nabla_\lambda \ln(\sqrt{g}\,\phi^4).
\eea
 From (\ref{tt1}), (\ref{t2}) 
\bea\label{nabla3}
\tilde\nabla_\lambda\, (\phi^2 g_{\mu\nu})
=(-2)\,\big(g_{\mu\nu}\,V_\lambda-g_{\mu\lambda} V_\nu - g_{\nu\lambda} V_\mu\big)\, \phi^2,
\eea

\medskip\noindent
so the theory is non-metric. From (\ref{nabla3})  we find the
 solution\footnote{Use that
$\tilde\nabla_\lambda g_{\mu\nu}=\partial_\lambda g_{\mu\nu}-\tGamma_{\mu\lambda}^\rho g_{\rho\nu}
-\tGamma_{\nu\lambda}^\rho \, g_{\mu\rho}$,  for cyclic permutations of indices and combine them.
} 
 $\tGamma$ to (\ref{eq23}) {\it in terms  of} $V_\lambda$:
\bea\label{Vm1}
\tGamma_{\mu\nu}^\alpha&=&\Gamma_{\mu\nu}^\alpha(\phi^2 g)
-\,\big(3\, g_{\mu\nu}\, V_\lambda
-g_{\nu\lambda}\,V_\mu - g_{\lambda\mu} \,V_\nu\,\big)\,g^{\lambda\alpha},
\\
\quad \rm{with} && \Gamma^\alpha_{\mu\nu}(\phi^2 g)=\Gamma^\alpha_{\mu\nu}(g)+1/2\, \big( \delta_\nu^\alpha
\, \partial_\mu +\delta_\mu^\alpha\, \partial_\nu -g^{\alpha\lambda} g_{\mu\nu} \,\partial_\lambda)
\ln\phi^2.\nonumber
\eea

\medskip\noindent
 $\Gamma_{\mu\nu}^\alpha(g)$ is  Levi-Civita connection of $g_{\mu\nu}$.
From (\ref{Vm1}),
$\tGamma_\lambda\!=\Gamma_\lambda(\phi^2 g)+ 2  V_\lambda$ and 
with (\ref{fmunu}), (\ref{t2}) 
\medskip
\bea\label{w}
\w_\lambda=-(1/2)\, \tilde\nabla_\lambda \ln\sqg,
\eea

\medskip\noindent
and finally,  $V_\lambda\!=\! \w_\lambda-\partial_\lambda\ln\phi^2$.
With this relation between $V_\lambda$ and $\w_\lambda$, 
 the solution $\tGamma$ in  (\ref{Vm1})  is finally
expressed as a function of $\w_\lambda$, $\phi$,  and  will be used shortly
 to compute the action for $\tGamma$ onshell (see eq.(\ref{proca1}) 
below\footnote{
As a remark, recall that eqs.(\ref{eom1}) for $\tGamma$ were shown to
 be equivalent to the  combined set of (\ref{eq23}), (\ref{t0}) and 
we solved (\ref{eq23}) with solution
  $\tGamma_{\mu\nu}^\alpha$ in (\ref{Vm1})  {\it expressed in terms
of} $V_\mu\sim \tGamma_\mu$. We  have four  remaining  equations in
(\ref{t0}) for  $\tGamma_\mu$ itself, that could in principle be used to also  ``fix''
 $\tilde\Gamma_\mu\sim V_\mu$  or equivalently 
$\w_\mu$, since  $V_\mu=\w_\mu-\partial_\mu\ln\phi^2$;  however
we do not do this step since  $\w_\mu$ is a {\it massless,  dynamical} gauge field
 enforcing  gauged scale symmetry  (\ref{s2}) 
 of initial  action (\ref{qq1}), (\ref{sss}). Then what information does (\ref{t0}) bring?
With eq.(\ref{tt1})  for $\lambda\!=\!\nu$, eq.(\ref{t0}) is actually
$\tilde\nabla_\rho F^{\rho\mu}+\alpha^2\xi_0 \phi^2 V^\mu=0$, 
which  is just  the equation of motion of  $\w_\lambda$; this may be seen from
final  Lagrangian (\ref{proca1})  which has all  $\tGamma_{\mu\nu}^\alpha$ onshell
(expressed in terms of $\tGamma_\mu$) but $\tGamma_\mu$ is  kept offshell,
for the reason mentioned; $\w_\lambda$ may be integrated out after 
 becoming massive, see  later.}).
Notice that solution  $\tilde\Gamma$ of (\ref{Vm1}) and also (\ref{nabla3}), are 
 invariant under transformations (\ref{cr}), (\ref{s2}) for any $\Omega(x)$
since $\phi^2 g_{\mu\nu}$, $V_\lambda$,  $\sqg \phi^4$ are invariant.

As expected,   $\w_\lambda$  is  the Weyl field of  non-metricity defined as 
$Q_{\lambda\mu\nu}\!\equiv \!\tilde\nabla_\lambda g_{\mu\nu}$, since  from (\ref{w}) the trace 
$Q^\mu_{\lambda\mu}=-4\, v_\lambda$.  Non-metricity  is a consequence of the dynamical $\w_\lambda$,
see (\ref{t0}).
Eq.(\ref{w})  is similar to that  in Weyl quadratic gravity of same  symmetry
(e.g.\cite{Fulton}).

Finally, from  solution (\ref{Vm1}) and (\ref{e02}) we compute $R_{\mu\nu}(\tGamma)$ and 
 scalar curvature\footnote{ $R_{\mu\nu}(\tGamma)$ has the following expression
 (which by contraction with $g^{\mu\nu}$  gives $R(\tGamma,g)$ of (\ref{s4})):
\bea\label{rr}
R_{\mu\nu}(\tGamma)&=& R_{\mu\nu}(g)- 3 g_{\mu\nu}
\big(\nabla^\lambda V_\lambda+V^\lambda \partial_\lambda\theta\big)
- \,(\nabla_\mu V_\nu -\nabla_\nu V_\mu)
-6 \, V_\mu V_\nu
+1/2\,(\partial_\mu\theta)(\partial_\nu\theta)
\nonumber\\[-7pt]
 &-&
1/2\, g_{\mu\nu} g^{\alpha\beta} (\partial_\alpha\theta)
 (\partial_\beta\theta)
- 1/2\,g_{\mu\nu}\nabla^\lambda\partial_\lambda\theta +1/2\,\nabla_\nu 
\partial_\mu\theta-3/2\,\nabla_\mu\partial_\nu\theta,\quad 
\theta\equiv \ln\phi^2.
\eea
\vspace{-0.33cm}}
$R(\tGamma,g)$
\medskip
\bea\label{s4}
R(\tGamma,g)=R(g)-6 g^{\mu\nu} \nabla_\mu\nabla_\nu\ln\phi
-6 (\nabla_\mu\ln\phi)^2
-12\, \big(\nabla_\lambda V^\lambda
+ V^\lambda \partial_\lambda\ln\phi^{{2}}\big)
-6  V_\mu\, V^\mu.
\eea

\medskip\noindent
$R(g)$ is here the usual  Ricci scalar and  $V_\lambda= w_\lambda-\partial_\lambda\ln\phi^2$.
Using (\ref{s4}) in (\ref{sss}), then finally
\medskip
\bea\label{proca1}
L_2= \sqrt{g}\,\Big\{ -\frac{\xi_0}{12}
\Big[\phi^2 R(g) +6 (\partial_\mu\phi)^2\Big]
+\frac{\xi_0}{2} \phi^2 \,(\w_\mu-\partial_\mu\ln\phi^2)^2
-\frac{1}{4\, \q^2} F_{\mu\nu}^2-\frac{\xi_0}{4!}\,\phi^4\Big\}.
\eea

\medskip
This is the ``onshell''  Lagrangian of  EP quadratic gravity of eq.(\ref{qq1}) and
is gauged scale invariant.
$L_2$  is a second-order scalar-vector-tensor theory of gravity which 
 is  ghost-free according to \cite{gh}
for a torsion-free connection as here (this is also obvious from (\ref{EP}) below).
This is relevant since  initial action  (\ref{qq1}) which (offshell)
was  of second  order  is actually a  four-derivative theory in the metric\footnote{
This agrees with e.g.\cite{SO1} that in general in a Palatini model 
its metric part leads to a fourth order theory.}
 for   $\tGamma$  onshell; indeed, $R(\tGamma,g)^2$ 
 in (\ref{qq1}) with replacement (\ref{s4}) contains the higher derivative term 
$R^2(g)+...$; 
this four-derivative  theory has an equivalent second-order formulation
 with additional $\phi$, as shown in  eq.(\ref{proca1}). 
 Finally, if $\w_\mu=\partial_\mu\ln\phi^2$ (``pure gauge''),  
the model is Weyl integrable  and (\ref{proca1}) 
recovers~(\ref{rt}). 

Lagrangian (\ref{proca1})  (also initial (\ref{e1}))
is similar to that of Weyl quadratic gravity \cite{Ghilen1,Ghilen2}, up to 
a Weyl tensor-squared term not included here. However, unlike in Weyl theory, 
  here $\tGamma$  is $\phi$-dependent; also, in Weyl theory
 non-metricity follows from the underlying Weyl 
conformal geometry, while here it emerges after we determine $\tGamma$ from
 its equation of motion.

\subsection{Stueckelberg breaking  to Einstein-Proca action}
Given $L_2$ in (\ref{proca1}) with gauged scale symmetry we would like 
to ``fix the gauge''.
We choose the Einstein gauge obtained from (\ref{proca1})  by  transformations
(\ref{cr}), (\ref{s2})   of a special  $\Omega^2\!=\!\xi_0\phi^2/(6 M^2)$ 
fixing $\phi$ to a {\it constant} ($\langle\phi\rangle\not=0$). 
After\,removing the\,hats (\,$\hat{}$\,) on transformed $g$,$v_\mu$, $R$, we find
\bea\label{EP}
L_2=\sqg\, \Big\{-\frac12 M^2 R(g) +3\, M^2 \,\w_\mu \,\w_\nu\, g^{\mu\nu} -\frac{1}{4 \q^2}
F_{\mu\nu}^2-\frac{3}{2\xi_0} M^4 \,\Big\}.
\eea
This is the Einstein-Proca action for the gauge field $\w_\mu$ with a positive cosmological constant,
in which we  identified $M$ with the Planck scale ($M$) as seen from eq.(\ref{proca1})
\bea
M^2\equiv\xi_0\langle\phi\rangle^2/6.
\eea
The initial gauged scale invariance is broken by a gravitational Stueckelberg mechanism
\cite{S1,P1,P2}:  the massless $\phi$ is not part of the action anymore, but $\w_\mu$ has
become massive, after ``absorbing'' the derivative  $\partial_\mu(\ln\phi)$ 
 of the Stueckelberg field (dilaton) in eq.(\ref{proca1}).
Note that  $\partial_\mu(\ln\phi)$ is actually the Goldstone of special conformal symmetry - 
this Goldstone is not independent but is determined by the derivative of the dilaton \cite{J}.
The number of degrees  of freedom (dof) other than graviton
is conserved in going from (\ref{proca1}) to (\ref{EP}), as it should be
for spontaneous breaking: 
massless $\w_\mu$ and dynamical $\phi$ are replaced by massive $\w_\mu$ (dof=3).
The  mass of $\w_\mu$  is $m_\w^2\!=6 \q^2 M^2$ which is near Planck scale $M$
(unless one fine-tunes  $\q\!\ll\! 1$).

Using the same transformation $\Omega$,  from (\ref{nabla3})
\bea\label{non1}
\tilde\nabla_{\lambda} g_{\mu\nu}=(-2)(g_{\mu\nu} \w_\lambda-g_{\mu\lambda} \w_\nu -g_{\nu\lambda} \w_\mu).
\eea
This has a solution  $\tGamma$ that is immediate from 
(\ref{Vm1}) for $\phi$ constant and $V_\lambda$ replaced by $\w_\lambda$.
Finally, after the  massive field $\w_\mu$ decouples,
metricity is recovered below  $m_\w$, so
$\tilde\nabla_{\lambda} g_{\mu\nu}\!=\!0$ and $\tGamma\!=\!\Gamma(g)$. 
Briefly, Einstein action is a ``low energy''  limit of  Einstein-Palatini
 quadratic gravity, and the Planck scale $M\sim \langle\phi\rangle$ is a phase transition 
scale (up to coupling $\q$)\footnote{A special case:
consider 
 (\ref{sss}) with $\phi^2\!\!=\!6 M^2\!/\xi_0$=constant, i.e. a different initial
action with no symmetry!  then (\ref{s4}), (\ref{proca1}) simplify;
we still find (\ref{EP}) but there is no dynamical $\phi$ and thus no
 Stueckelberg mechanism.}.

 For comparison, in Weyl quadratic gravity e.g. \cite{Ghilen1,Ghilen2}, non-metricity is
 different\footnote{Contracting (\ref{non1}),(\ref{non2}) by  $g^{\mu\nu}\!$ gives the same
 non-metricity trace, justifying our normalization  of $\w_\mu$ eq(\ref{fmunu}).}
\bea\label{non2}
\tilde\nabla_\lambda g_{\mu\nu}=-g_{\mu\nu}\,\w_\lambda.
\eea
Interestingly the different non-metricity of these  theories (giving different $\tGamma$)
has phenomenological impact,
see Section~\ref{5}.  In  both theories the non-metricity scale is $m_\w\sim\,$Planck scale
and is large enough (current bounds \cite{Latorre,Lobo} are low $\sim$TeV) to 
suppresses unwanted effects e.g. atomic spectral lines spacing.
Past critiques of non-metricity   assumed  a massless $\w_\mu$.

Finally, let us  remark that the above spontaneous symmetry breaking mechanism
for initial action (\ref{qq1}) is special since it 
takes place  {\it in the absence of matter}. Indeed, the necessary scalar (Stueckelberg) field
$\ln\phi$
was not added ad-hoc to this purpose, as usually done in the literature; instead, this
field was ``extracted'' from the $R^2$ term in the initial, symmetric action (\ref{qq1}) and is
thus of  geometric origin. This situation is  similar to  Weyl quadratic gravity
where this mechanism  was first  noticed \cite{Ghilen1,Ghilen2}.

\subsection{Conserved current}\label{3.3}

Eqs.(\ref{t0}) and (\ref{tt1}) show there is now a non-trivial current 
due to dynamical $\w_\mu\sim\tGamma_\mu$
\medskip
\bea\label{consJ}
J^\mu\!
=\sqg\, g^{\rho\mu} \,\phi\, (\partial_\rho-1/2\,  v_\rho)\,\phi,\qquad
 \nabla_\mu J^\mu\!=\!0,
\eea 

\medskip\noindent
This  is conserved since
$F_{\mu\nu}$ in (\ref{t0}) is anti-symmetric. To obtain (\ref{consJ})
 we used that the lhs of (\ref{t0}) and of (\ref{tt1}) (with $\lambda\!=\!\nu$) are equal
and replaced  $V_\lambda=\w_\lambda-\partial_\lambda\ln\phi^2$.
The current  $J^\mu$ is the same as that in  Weyl quadratic gravity \cite{Ghilen2} (eq.18)
which has similar symmetry but different non-metricity.
The presence of this conserved current extends  to the case of the 
{\it gauged} scale symmetry a similar conservation 
for  a global scale symmetry  \cite{FRH1}.
For
a  Friedmann-Robertson-Walker metric with $\phi$ only $t$-dependent
 such current conservation in the global case
 naturally  leads to  $\phi$=constant \cite{FRH1}
 and  a breaking of  scale symmetry. 
In our case, since eq.(\ref{EP})  has $\phi$=constant (assumed $\langle\phi\rangle\not =0$), 
 then from (\ref{consJ}) one has $\nabla_\mu v^\mu=0$ which is  a condition
 similar to that for a Proca (massive) gauge field, leaving 3 degrees of freedom 
 for $\w_\mu$ in (\ref{EP}).

\bigskip
\section{Palatini quadratic gravity: adding matter}
\label{4}

In this section we re-do the previous analysis in the presence
of a scalar $\chi$ which   can be the SM Higgs,  with non-minimal coupling
 with {\it Palatini connection} to the EP quadratic gravity.

The general Lagrangian of the field 
$\chi$, with gauged scale invariance, eqs.(\ref{cr}),(\ref{s2}) is
\medskip
\bea\label{ll1}
L_3=\sg\,\Big[\,
\frac{\xi_0}{4!}\,R(\tGamma,g)^2 - \frac{1}{4 \q^2}\,
F_{\mu\nu}^2 - \frac{1}{12} \,\xi_1\chi^2\,R(\tGamma,g)
+\frac12 \,(\tilde D_\mu\chi)^2 - \frac{\lambda_1}{4!}\chi^4\Big],
\eea

\medskip\noindent
with the potential dictated by this symmetry and with
\medskip
\bea
\tilde D_\mu\chi=(\partial_\mu-1/2\,\w_\mu)\,\chi.
\eea

\medskip\noindent 
Under (\ref{cr}), (\ref{s2}) the  Weyl-covariant derivative transforms as
$\hat{\tilde D}_\mu\hat \chi=(1/\Omega)\, \tilde D_\mu\chi$.
As in previous sections,  replace $R(\tGamma,g)^2\!\ra\! -2 \phi^2 R(\tilde\Gamma,g)-\phi^4$
 to find an equivalent ``linearised''  $L_3$
\medskip
\be\label{ll1prime}
L_3=\sg\,\Big[\,
- \frac{1}{2} \,\rho^2\,R(\tGamma,g)
- \frac{1}{4 \q^2}\,F_{\mu\nu}^2 
+
\frac12 \,(\tilde D_\mu\chi)^2
-\cV(\chi,\rho)
\Big],
\ee

\medskip\noindent
where
\medskip
\bea\label{toE}
\cV(\chi,\rho)\equiv \frac{1}{4!}\,\Big[\,
\frac{1}{\xi_0} \big(6\rho^2-\xi_1\chi^2\big)^2
+
\lambda_1\chi^4\,\Big],\qquad\text{and}\qquad
\rho^2=\frac16\,\big(\xi_1\chi^2+\xi_0\phi^2).
\eea

\medskip\noindent
Notice that we also replaced the scalar field $\phi$ by the new,  radial direction field $\rho$;
$\ln\rho$ transforms as $\ln\rho\rightarrow \ln\rho -\ln\Omega$ and
acts as the (would-be) Goldstone  of the symmetry.

The equation of motion for $\tGamma_{\mu\nu}^\lambda$ is similar to (\ref{eom1})
 but with a replacement $\phi\ra\rho$  
and with an additional contribution from  the kinetic term of $\chi$. 
Following the same steps as in the previous section, we eliminate the 
contributions of the kinetic terms of $\chi$ and $\w_\mu$ to the equation of
$\tGamma$ and find an equation similar to (\ref{eq23}) with $\phi\ra \rho$: 
\medskip
\bea
\tilde\nabla_\lambda (\sqg g^{\mu\nu}\rho^2)
-\frac15 \Big\{
\delta_\lambda^\nu\,
\tilde\nabla_\sigma (\sqg g^{\sigma\mu}\rho^2)+(\mu\leftrightarrow\nu)
\Big\}=0.
\eea

\medskip\noindent
This gives (see previous section):
\bea\label{nabla2}
\tilde\nabla_\lambda (\rho^2 g_{\mu\nu})
=(-2)\rho^2 (g_{\mu\nu}\,V_\lambda-g_{\mu\lambda} V_\nu - g_{\nu\lambda} V_\mu),
\eea

\medskip\noindent
where $V_\mu=(-1/2)\tilde\nabla_\mu \ln (\sqg \rho^4)=\w_\mu-\partial_\mu\ln\rho^2$.
From (\ref{nabla2}) one finds the solution for Palatini connection $\tGamma_{\mu\nu}^\alpha$  in 
terms of $\w_\mu\sim\tGamma_\mu$, with a result  similar to (\ref{Vm1}) but with 
 $\phi\ra \rho$.  We use this solution for the connection back in the action and
 find for $\tGamma$ onshell\footnote{
As in Section~\ref{3}, the trace  $\tGamma_\mu\sim \w_\mu$  is kept offshell since
 we do not integrate out  {\it massless} dynamical $\w_\mu$.}
\medskip
\be
L_3=\sqg \Big\{
\frac{-1}{2} \Big[\rho^2 R(g) + 6 (\partial_\mu\rho)^2\Big] + 3 \rho^2 (\w_\mu-\partial_\mu\ln\rho^2)^2
-\frac{1}{4 \q^2} \,F_{\mu\nu}^2 +\frac12 (\tilde D_\mu\chi)^2 - \cV(\chi,\rho)\Big\}.
\ee

\medskip\noindent
$L_3$ has a gauged scale symmetry and extents (\ref{proca1}) in the presence of scalar field $\chi$.

Finally, we choose  the Einstein gauge by using  transformation
(\ref{cr}),(\ref{s2}) of a particular $\Omega\!=\!\rho/M$ which essentially
sets $\hat\rho$ to a {\it constant} (vev). In terms of the  new variables (with a hat) we find
\medskip
\be\label{W3}
L_3=
\sqrt{\hat g}\, \Big\{-\frac{1}{2} M^2\,R(\gh) +3  M^2 \hat \w_\mu\hat \w^\mu
-\frac{1}{4 \q^2}  \hat F_{\mu\nu}^2 +\frac{1}{2}(\hat{\tilde D}_\mu\hat\chi)^2
-\cV(\hat\chi,M) \Big]\Big\},
\ee

\medskip\noindent
with ${\hat{\tilde D}}_\mu\hat\chi=(\partial_\mu-1/2\,\,\hat \w_\mu )\hat\chi$
and we identify $M$  with  the Planck scale ($M=\langle\hat\rho\rangle$).
As in the absence of  matter, we obtained the Einstein-Proca action of a  gauge field 
that became massive after  Stueckelberg  mechanism  of ``absorbing'' the  
derivative term  $\partial_\mu(\ln\rho)$.  A canonical kinetic term of $\hat\chi$ 
remained present in the action,
 since  only one  degree of freedom (radial direction $\rho$) was  ``eaten'' by $\w_\mu$.
The mass of $\w_\mu$ is  $m_\w^2=6 \q^2 M^2$.
The potential becomes
\medskip
\bea\label{fff}
\cV= \frac{3 M^4}{2\,\xi_0} 
\Big[1-\frac{\xi_1\hat\chi^2}{6 \,M^2}
\Big]^2
+
\frac{\lambda_1}{4!}\,\hat\chi^4.
\eea

\medskip\noindent
For  a  ``standard''  kinetic term for $\hat\chi$,
similar to a  ``unitary  gauge'' in electroweak case, we   remove the
coupling  $\hat \w^\mu \partial_\mu \hat\chi$ in the Weyl-covariant derivative in (\ref{W3})
by a field redefinition 
\medskip
\bea\label{tt}
{\hat \w_\mu^\prime} =\hat \w_\mu - \partial_\mu \ln \cosh^2 \Big[
\frac{\h}{2 M\sqrt{6}}\Big],\qquad
\hat \chi=2 M\sqrt{6}\,\sinh\Big[\frac{\h}{2 M\sqrt 6}\Big],
\eea

\medskip\noindent
which replaces $\hat\chi\ra\sigma$. After some algebra, we find the final Lagrangian
\medskip
\be\label{twp}
L_3=\sqrt{\hat g}\, \Big\{
-\frac{1}{2} M^2\,\hat{R}
+ 3  M^2 \cosh^2 \Big[\frac{\h}{2 M\sqrt 6}\Big] 
\,\hat \w^\prime_\mu\hat \w^{\prime\,\mu}
-\frac{1}{4 \q^2} \hat F^{\prime\, 2}_{\mu\nu}  
+\frac{\hat g^{\mu\nu}}{2}\partial_\mu\h \partial_\nu \h
-{\hat \cV}(\h)\Big\}
\ee
with
\bea\label{scalars2}
{\hat \cV(\h)}=
\hat\cV_0\,
\Big\{
 \Big[
1-4 \xi_1\sinh^2 \frac{\h}{2 M\sqrt 6}\Big]^2+
16 \,\lambda_1  \xi_0 \sinh^4\frac{\h}{2 M\sqrt 6}
\,\Big\},\quad 
\hat\cV_0\equiv\frac{3}{2} \frac{M^4}{\xi_0}.
\label{ov1}
\eea

\medskip\noindent
In (\ref{twp}) one finally rescales $\hat \w^\prime_\mu\ra \q\,\hat \w_\mu^\prime$ 
for a canonical gauge kinetic term.

For {\it small} field values,  $\h\ll M$, then $\hat\chi\approx \h$ (up to $\cO(\h^3/M^2)$)
 and a SM Higgs-like potential is recovered\footnote{We shall see shortly that inflation
``prefers'' ultraweak or vanishing values for $\lambda_1$ in (\ref{toE}) and (\ref{fff}).},  
see eq.(\ref{fff}).
For $\xi_1>0$ it has spontaneous breaking of the symmetry carried by $\h$ 
i.e.  electroweak  (EW) symmetry if $\h$ is  the Higgs; 
this is  triggered by the non-minimal coupling to gravity 
($\xi_1\!\not=\!0$) and Stueckelberg mechanism. 
The negative mass term originates in (\ref{toE}) due to the  $\phi^4$ term
(itself induced by $\tilde R^2$). The mass  $m_\h^2\!\propto\! \xi_1 M^2/\xi_0$
may  be small enough, near the EW scale  by tuning  $\xi_1\!\ll\!\xi_0$.
It may be interesting to
study if the gauged scale symmetry brings some ``protection''  to $m_\h$ at 
the quantum level.

$L_3$ of (\ref{twp})  is similar to that in Weyl quadratic gravity with a non-minimally 
coupled scalar/Higgs field
 \cite{Ghilen1,Ghilen2,Winflation}\footnote{For comparison to Weyl gravity Lagrangian
see e.g. eqs.(39)-(41) in \cite{Ghilen2} and eqs.(21), (22) in \cite{Ghilen1}.},
 up to a rescaling of 
the couplings ($\xi_1$, $\lambda_1$) and fields ($\h$).
This difference is due  to the different non-metricity of the two theories,
 eqs.(\ref{non1}), (\ref{non2}). Both cases provide a gauged scale invariant theory of
 quadratic gravity coupled to matter. They both recover Einstein gravity in their
 broken phase, see eq.(\ref{twp}), and also  metricity below the scale $m_\w\sim \q M$ ($\q\leq 1$).
This result may be more general - it may apply to other theories with this symmetry and
can be used for model building.

To conclude,
mass generation (Planck scale, $\w_\mu$ mass) and Einstein gravity emerge naturally from
spontaneous breaking of gauged scale symmetry in Einstein-Palatini theories,
even in the absence of matter.
Actions (\ref{qq1}), (\ref{ll1}) were  inspired by Weyl quadratic
gravity of similar breaking  \cite{Ghilen1};
but in a more general case, additional operators 
may be present in  (\ref{qq1}), (\ref{ll1});  for a  list of all
quadratic operators and a complementary  study see \cite{c6}. 
The mechanism of symmetry breaking should remain valid in their presence if one includes the terms in (\ref{qq1}):
$R^2$ that 'supplied' the scalar field and $R_{[\mu\nu]}^2$ generating the symmetry and
non-metricity. However, in  such general  case it is unclear that one can
still solve  algebraically 
the second-order differential equations  of motion of $\tGamma$ (eqs.(\ref{eom1}))
without  simplifying assumptions, since
these  equations acquire new terms of different indices structure and  new states will
be present (ghosts, etc).

\section{Palatini $R^2$ inflation}\label{5}

In this section we consider an application to inflation of 
the action in the previous section.

For {\it large} field values, 
the  potential in (\ref{scalars2}) can also be used for inflation 
(hereafter  Palatini $R^2$ inflation),  with $\sigma$ as the inflaton\footnote{Unlike
in Starobinsky models, there is no scalaron here, its counterpart was ``eaten'' by massive $\w_\mu$. 
}.
For a Friedmann-Robertson-Walker metric (FRW)  $(1,-a^2(t),-a^2(t), -a^2(t))$
and compatible background $\w_\mu(t)=(\w_0(t),0,0,0)$ the gauge fixing condition $\nabla_\mu \w^\mu=0$
gives that $\w_\mu(t)$ redshifts to zero $\w_\mu(t)\sim 1/a^3(t)$. Then
the coupling $\w_\mu-\sigma$ in (\ref{twp}) is vanishing and therefore
$\w_\mu(t)$ cannot affect inflation; this means we have
single-field inflation of  potential (\ref{scalars2})  and standard slow-roll formulae can be used.
Further, since $M$ is just a phase transition scale, field values $\h\!\geq\! M$ are natural.
$\hat\cV(\h)$  is similar to that in Weyl gravity $R^2$-inflation, see 
\cite{Winflation,Ross} for a  detailed analysis\footnote{For 
 related works on inflation in the Einstein-Palatini formalism see e.g.
\cite{II1,II2,II3,II4,II5,II6,II7,Palatini1,Palatini2,Palatini3}.}; however,
 as mentioned, the couplings and field normalization in the potential differ 
(for same initial couplings and non-metricity trace);
 hence the  spectral index $n_s$ and  tensor-to-scalar ratio $r$ are different, too,
and need to be analyzed separately.

\begin{figure}[t!]
\begin{center}
\includegraphics[height=0.42\textwidth]{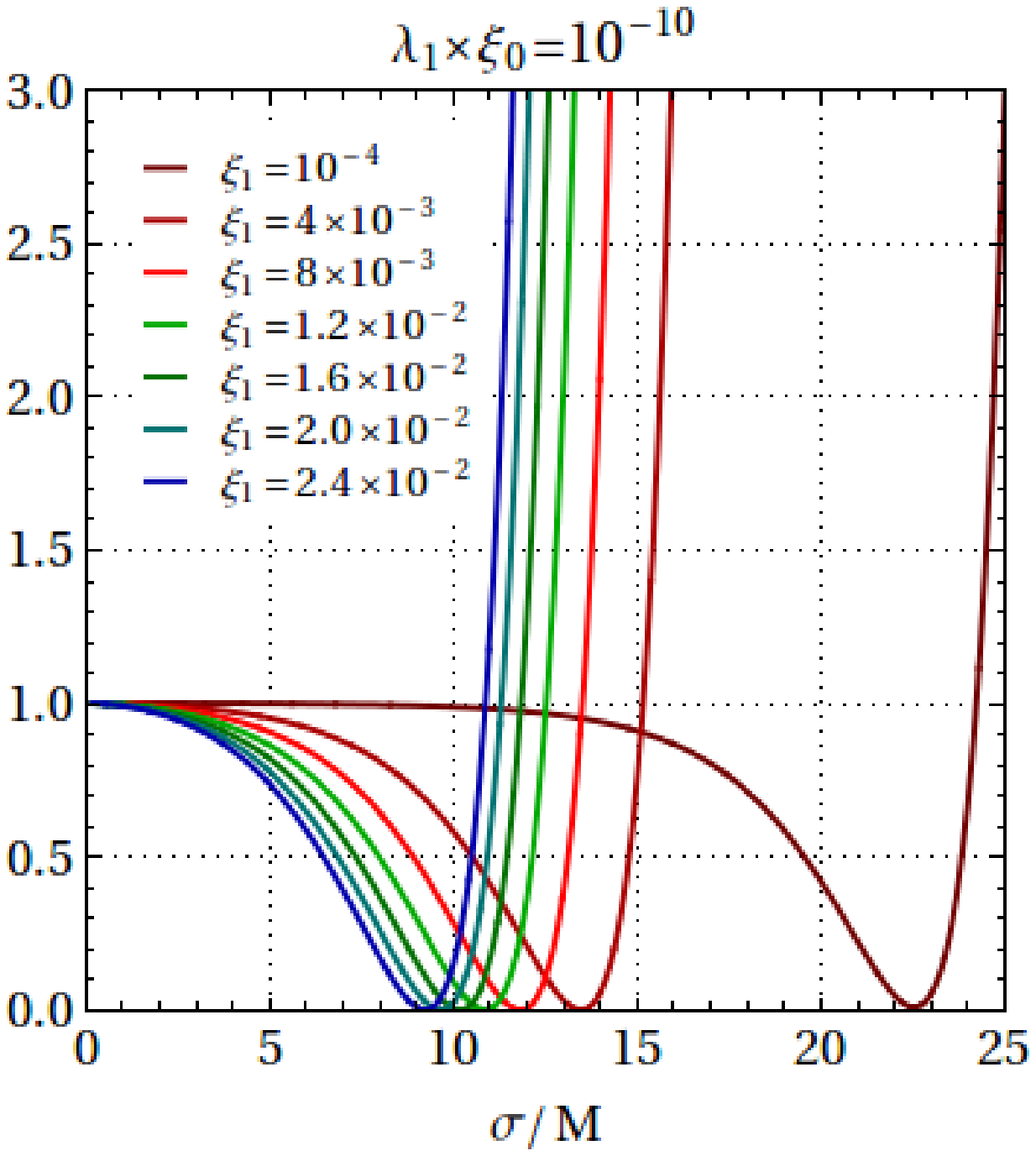}
\includegraphics[height=0.42\textwidth]{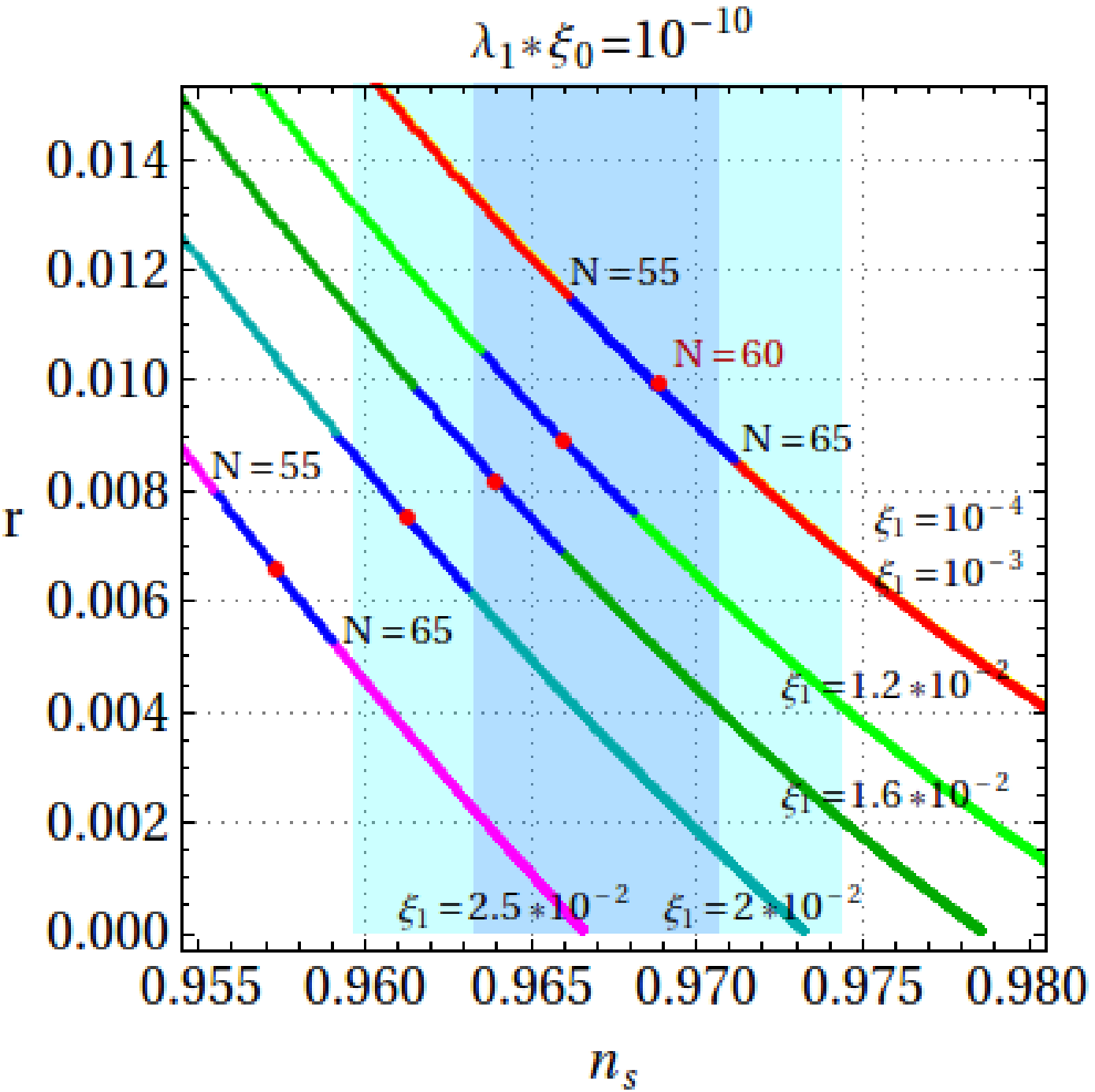}
\end{center}
\vspace{-0.2cm}
\caption{\small
{\bf Left plot:} The potential $\hat\cV(\sigma)/\hat\cV_0$ for  $\lambda_1\xi_0=10^{-10}\!\ll\! \xi_1^2$ 
with different   $\xi_1\!\ll\! 1$.  For larger $\lambda_1\xi_0$ the curves move to the left while 
the minimum of the rightmost ones is lifted.
Larger values of $\lambda_1\xi_0$ are allowed, but inflation becomes less likely when 
$\lambda_1\xi_0\sim \xi_1^2$.  
The flat region is wide for a large range of $\h$, with the width controlled  by
$1/\sqrt\xi_1$ while its height is $\hat\cV_0\propto 1/\xi_0$. 
We have  $\hat\cV/\hat\cV_\text{min}\propto\xi_1^2/(\lambda_1\xi_0)$.
{\bf Right plot:} The values of $(n_s,r)$ for different values of $\xi_1$ that enable values of 
$n_s=0.9670\pm 0.0037$ at 68\% CL (blue band) and 95\% CL (light blue region).
For each curve $N=60$ efolds is marked by a red point and the dark blue interval corresponds to
$55\leq N\leq 65$. Curves of  $\xi_1\!<\!10^{-3}$  are degenerate with the red one
while those with $\xi_1\!>\!2.5\! \times\! 10^{-2}$ have $N\!>\!65$.
}\label{fig1}
\end{figure}

The potential is shown in Figure~\ref{fig1} for perturbative values of the couplings relevant 
for successful inflation. This demands 
 $\lambda_1 \xi_0\ll \xi_1^2\ll 1$, with the
 first relation  from demanding that the initial energy be larger than
 at the end of inflation $\hat\cV_0>\hat\cV_\text{min}$, respected by 
choosing a small enough $\lambda_1$ for given $\xi_{0,1}$. 
Therefore, we shall  work in the leading order in $(\lambda_1\xi_0)$.

The slow-roll parameters are:
\bea\label{eqeps}
\epsilon&=&
\frac{M^2}{2} \Big\{\frac{\hat\cV^\prime(\h)}{\hat\cV(\h)}\Big\}^2
= 
\frac{{4}}{3}\, \,\xi_1^2\, \sinh^2 \frac{\h}{M\sqrt 6} +\cO(\xi_1^3)\,
\\
\eta&=&
M^2\,\frac{\hat\cV^{\prime\prime}(\h)}{\hat\cV(\h)}=
-\frac23\,\xi_1 \,\cosh {\frac{\h}{M\sqrt 6}}+\cO(\xi_1^2)\,
\label{eqeta}
\eea
Then
\bea\label{nsns}
n_s\, =\, 1 + 2\,\eta_* - 6\, \epsilon_*=
1 -\frac43\,\xi_1 \,\cosh\frac{\h_*}{M\sqrt 6}+\cO(\xi_1^2),\,
\eea

%
%
\noindent
where $\h_*$ is the value of $\h$ at  the horizon exit.
 With $r=16 \epsilon_*$  we have\footnote{
There is  also a  constraint on the parametric space
from the normalization of CMB anisotropy
 $\cV_0/(24 \pi^2 M^4 \epsilon_*)=\kappa_0$, $\kappa_0=2.1\times 10^{-9}$
and $r=16 \epsilon_*$ with $r<0.07$ \cite{planck2018} then $\xi_0=1/(\pi^2 r \kappa)
\geq 6.89 \times 10^8$. 
The aforementioned condition $\lambda_1 \xi_0\!\ll \!\xi_1^2$ is then
respected for perturbative $\xi_1$, $1/\xi_0$
  by choosing small $\lambda_1\!\ll\! \xi_1^2/\xi_0$.
}
\smallskip
\bea\label{rns}
r={12}\, (1- n_s)^2+\cO(\xi_1^2)
\eea
The contribution of $\epsilon$ is subleading for small $\xi_1$ considered here.
The slope of the curves in the plane $(n_s,r)$, shown in leading order in (\ref{rns}),
 is steeper than in Weyl $R^2$ inflation  \cite{Winflation} (or Starobinsky model)
where $r=3 (1-n_s)^2+\cO(\xi_1^2)$.

The exact numerical results  for $(n_s,r)$ in our model, for different  e-folds number $N$, are 
shown in  Figure~\ref{fig1}.
From experimental data   $n_s=0.9670\pm 0.0037$  ($68\%$ CL) and $r<0.07$ ($95\%$ CL) 
from Planck  2018 (TT, TE, EE + low E + lensing + BK14 + BAO) \cite{planck2018}.
Using this data, Figure~\ref{fig1} (right plot) shows that  a specific, small  range for $r$ is 
predicted in our model for the current range for $n_s$ at $95\%$ CL:
\bea\label{paa}
 N=60,\qquad  0.007 \leq r\leq 0.010,\qquad \textrm{[Palatini\,\,$R^2$\,\,inflation]}.
 \eea
Similar values for $r$ can be read from Figure~\ref{fig1} for 
 $55\leq N\leq 65$. The lower bound on $r$ comes from that for $n_s$
while the upper one corresponds to a  saturation limit, $\xi_1\!\ra\!0$, with values 
$\xi_1\!<\!10^{-3}$ having similar $(n_s,r)$. 
One should also respect the constraint  $\lambda_1\leq \xi_1^2/\xi_0$, 
 giving  $\lambda_1\sim 10^{-12}$ or smaller
(with the CMB anisotropy constraint $\xi_0\!\geq\! 6.89\times 10^8$).

For comparison, in Weyl $R^2$-inflation for 
same  $n_s$ at $95\%$ CL one has a 
smaller $r$  \cite{Winflation,Ross}
\bea\label{wey}
N=60, \qquad 0.00257\leq r\leq 0.00303,\qquad \textrm{[Weyl\,\,$R^2$\,\,inflation]}.
\eea

\medskip\noindent
The different  range for $r$ in eq.(\ref{paa}) versus eq.(\ref{wey}) 
is important since it enables us to 
 distinguish  these two inflation models  based on gauged scale invariance,
 and is due to their different non-metricity\footnote{
 For a more detailed comparison of Einstein-Palatini $R^2$-inflation to Weyl $R^2$-inflation
see \cite{toc}.}$^,$\footnote{
In the Starobinsky model  \cite{Sta} for similar $n_s$ one has $r\sim \cO(10^{-3})$, e.g. 
$r=0.0034$ ($N=55$) \cite{Patrignani:2016xqp}.}.
Such values for $r\sim 10^{-3}$ will soon be reached by various
 CMB experiments   \cite{CMB1,CMB2,CMB3}
that will then be able  test both  models. 
This  establishes an interesting connection between  non-metricity 
and testable  inflation predictions.

Similar values for $r$ were found in other recent inflation models in Palatini $R^2$ gravity 
\cite{Palatini1,Palatini2,Palatini3} but these are not gauged scale invariant.
In the absence of this symmetry, other successful models (e.g. Starobinsky model \cite{Sta})
 have  corrections to $r$ from higher curvature operators ($R^4$, etc) of unknown coefficients 
\cite{Edholm}.
Such operators (and their corrections) are not allowed here because 
they   must  be suppressed by some effective scale whose presence would  violate 
 scale invariance\footnote{The dilaton field cannot suppress
them itself since  it  is ``eaten'' to all orders by the 
gauge field $\w_\mu$.}.  Another advantage is that due to the {\it gauged} scale symmetry
 Palatini $R^2$ inflation is allowed by black-hole physics (similarly for Weyl 
$R^2$ inflation \cite{Winflation}), 
 in contrast to  models of  inflation with {\it global} scale symmetry\footnote{A global  
symmetry is broken since global charges can be eaten by black holes which then evaporate 
\cite{BH}.}.

\section{Conclusions}\label{6}

At a fundamental level gravity may be regarded as a theory of connections. An example 
is  the Einstein-Palatini (EP) approach to gravity where the  connection ($\tGamma$) is apriori 
independent of the metric, and is determined by its  equation of motion, from the action.
 For simple actions $\tGamma$ plays an auxiliary role (no dynamics) and can be solved algebraically. 
In particular, for
Einstein action in the EP approach one finds that the connection is actually equal to the 
Levi-Civita connection (of the metric formulation); then
Einstein gravity is recovered, so the metric and EP approaches  are  equivalent.
However, this equivalence is not true in general, for complicated actions, etc.
 In this work we  considered quadratic gravity actions in the EP approach, with the goal to show that,
 while  this equivalence does not hold true,  one can still find  actions  that recover
 {\it dynamically} the Levi-Civita connection, metricity,  Einstein gravity and Planck mass
in  some ``low-energy'' limit, even in the {\it absence} of matter.

We studied  EP quadratic gravity given by
  $R(\tGamma,g)^2+R_{[\mu\nu]}(\tGamma)^2$ which has local scale symmetry.
 $R_{[\mu\nu]}(\tGamma)^2$ can be regarded
  as a  gauge kinetic term for the vector field $\w_\mu\sim \tGamma_\mu -\Gamma_\mu$ 
where $\tGamma_\mu$ ($\Gamma_\mu$) denotes the trace of the  Palatini  (Levi-Civita) connections,
respectively. Hence this theory actually  has a {\it gauged}
 scale symmetry, with $\w_\mu$  the Weyl gauge field. 
 A consequence  of this symmetry is that  the theory is non-metric i.e.
$\tilde\nabla_\mu g_{\alpha\beta}\not=0$ (due  to  dynamical $\w_\mu \sim \tGamma_\mu$). 
 At the same time,  the  equations of motion of the connection ($\tGamma$) become  
complicated second-order differential equations and we showed how to solve them {\it algebraically}
in terms of an auxiliary scalar $\phi$ that ``linearises'' the  $R(\tGamma,g)^2$ term.
While initially  the action appears to be  of second order, 
for $\tGamma$ onshell
 it is a {\it higher derivative} theory since  $R(\tGamma,g)^2$   contains  a 
(four-derivative) metric contribution  $R(g)^2+...$. We showed that for $\tGamma$ 
onshell,  the action is equivalent to a second-order theory
in which the initial auxiliary field $\phi$ has become dynamical,
while preserving the symmetry of the theory.

The main result is that our EP quadratic gravity action 
has  an elegant spontaneous breaking mechanism of gauged scale invariance
and mass generation valid even in the absence of matter;
in this, the necessary scalar field ($\phi$) was not added ad-hoc
 to this purpose (as usually done), but was ``extracted'' from the $R^2$ term, as
mentioned, being of geometric origin.
The derivative $\partial_\mu\ln\phi$  of this field acting as a Stueckelberg field
is ``eaten'' by $\w_\mu$ which becomes  massive, of mass $m_\w$
proportional to the Planck scale $M\!\sim\!\langle\phi\rangle$.
  One obtains  the Einstein-Proca action for the gauge field $\w_\mu$ and a positive
cosmological constant. This is a ``low-energy'' broken phase of the initial action.
Below the scale $m_v\!\sim\! M$, the Proca field $\w_\mu$ decouples and  metricity and the Einstein 
action are recovered.  Non-metricity effects are strongly suppressed by a
 large scale ($\propto\! M$), which is important for the theory to be viable.

The above results remain valid in the presence of scalar matter (Higgs,  etc) 
with a (perturbative) non-minimal coupling  to this theory with a {\it Palatini 
connection};  in such case
and following the Stueckelberg mechanism, the scalar potential also has
a breaking of the symmetry under which this scalar  is charged,
e.g. electroweak symmetry in the higgs case.  
This is relevant for building models with this symmetry  for physics beyond the SM.

To summarise, Einstein-Palatini quadratic gravity  $R(\tGamma,g)^2+R_{[\mu\nu]}^2(\tGamma)$ 
is a gauged theory of scale invariance that is  spontaneously broken to 
the Einstein-Proca action for the Weyl field with a  positive cosmological constant;
if initial action also contains (non-minimally coupled) scalar fields  with Palatini connection,
a scalar potential is also present.

This picture  is similar  to a recent analysis  for  the original  
Weyl quadratic gravity,  despite the different non-metricity of these two theories.
With hindsight, this is not too surprising, since in both theories there is a gauged scale symmetry 
and the connection is not fixed by the metric,  except that in Weyl gravity  non-metricity is present 
from the onset (due to underlying Weyl conformal geometry) while here it 
emerges  for $\tGamma$ onshell. It is worth studying further 
the relation of these two theories, by including any  remaining operators 
(on the Einstein-Palatini side) that can have this symmetry.

There are also interesting predictions from inflation. 
While the scalar  potential is Higgs-like for small field values ($\ll\!M$),
 for large field values it can be used for inflation. With  the Planck scale $M$
a simple  phase transition scale, field  values above $M$ are  natural.
The inflaton potential is similar to that in  Weyl quadratic gravity, up to couplings and
field redefinitions  (due to different non-metricity of the two theories).
We find a specific  prediction  for the tensor-to-scalar ratio, $0.007\!\leq\! r\! \leq\! 0.01$,
for the current value of the  spectral index at $95\%$ CL.
This value of  $r$ is mildly larger than that predicted by  inflation in Weyl gravity.
This  enables us to distinguish   and test these two theories
by future CMB experiments that will reach such values of $r$. 
It also establishes an interesting connection between
non-metricity and  inflation predictions.

\bigskip\noindent
{\bf Acknowledgements:} The author thanks Graham Ross for helpful discussions on this topic
 at an early stage of this  work.

\vspace{1cm}
\section*{Appendix}

For a self-contained presentation, we include here  basic
aspects of the Palatini formalism used in the text.
In the (pseudo-)Riemannian geometry, the  Levi-Civita connection $\Gamma(g)$
is determined by the metric. In general, however, the connection can be introduced
 without reference to $g_{\mu\nu}$.
In the Palatini approach 
the connection $\tGamma$ is apriori independent of the metric
and is  determined  by the equations of motion.
To  ensure the covariant derivatives transform under  coordinate 
change  ($x\ra x'(x)$) as true  tensors, Palatini connection 
has a  transformation law
\bea
\tGamma^{ ' \lambda}_{\mu\nu}=
\frac{\partial x^{\alpha}}{\partial x^{' \mu}}
\frac{\partial x^{\beta}}{\partial x^{' \nu}}\,\,
\frac{\partial x^{' \lambda}}{\partial x^{\rho}}\,
\tGamma^{\rho}_{\alpha\beta}
+
\frac{\partial x^{' \lambda}}{\partial x^\sigma}
\frac{\partial^2 x^\sigma}{\partial x^{' \mu} \partial x^{' \nu}}
\eea
with $\tGamma'=\tGamma'(x')$, $\tGamma=\tGamma(x)$. In the text we also assumed 
$\tGamma_{\mu\nu}^\rho=\tGamma_{\nu\mu}^\rho$ (no torsion).
The Levi-Civita connection $\Gamma_{\mu\nu}^\lambda(g)$ has a similar transformation
\bea
\Gamma^{ ' \lambda}_{\mu\nu}(g)=
\frac{\partial x^{\alpha}}{\partial x^{' \mu}}
\frac{\partial x^{\beta}}{\partial x^{' \nu}}\,\,
\frac{\partial x^{' \lambda}}{\partial x^{\rho}}\,
\Gamma^{\rho}_{\alpha\beta}(g)
+
\frac{\partial x^{' \lambda}}{\partial x^\sigma}
\frac{\partial^2 x^\sigma}{\partial x^{' \mu} \partial x^{' \nu}}.
\eea

\medskip\noindent
Note that the difference of these connections transforms  as a tensor
\medskip
\bea
\tGamma^{ ' \lambda}_{\mu\nu}-\Gamma^{ ' \lambda}_{\mu\nu}(g)=
\frac{\partial x^{\alpha}}{\partial x^{' \mu}}
\frac{\partial x^{\beta}}{\partial x^{' \nu}}\,\,
\frac{\partial x^{' \lambda}}{\partial x^{\rho}}\,
(\tGamma^{\rho}_{\alpha\beta}(x)
-\Gamma^{\rho}_{\alpha\beta}(g)).
\eea

\medskip\noindent
Setting $\lambda=\nu$ and with the notation $\tGamma_\mu\equiv \tGamma_{\mu\nu}^\nu$, 
$\Gamma_\mu\equiv \Gamma_{\mu\nu}^\nu$, etc, then
\bea
\tGamma_\mu' - \Gamma_\mu' (g)=
\frac{\partial x^{\alpha}}{\partial x^{' \mu}}\,
(\tGamma_\alpha-\Gamma_\alpha(g))
\eea

\medskip\noindent
and  therefore $\w_\mu$ introduced in Section~\ref{3}
transforms as a covariant vector
\bea
\w_\mu'(x')=
\frac{\partial x^{\alpha}}{\partial x^{' \mu}}
\w_\alpha(x)
\eea

\medskip\noindent
Further,  the covariant derivatives used in the text are 
\medskip
\bea
\tilde\nabla_\nu \w_\mu=\partial_\nu \w_\mu-\tGamma^\lambda_{\mu\nu} \w_\lambda,
\qquad
\tilde\nabla_\nu \w^\mu=\partial_\nu \w^\mu+\tGamma^\mu_{\lambda\nu} \w^\lambda.
\eea
One also has
$\tilde \nabla_\lambda g_{\mu\nu}=\partial_\lambda g_{\mu\nu} 
-\tGamma^\rho_{\mu\lambda} g_{\rho\nu} -\tGamma^\rho_{\nu\lambda} g_{\rho\mu}$, 
also used in the text.

In Section~\ref{3} we introduced the gauge field $\w_\mu$ in  eq.(\ref{fmunu}).
This is general. For example, in Weyl gravity 
of similar gauged scale symmetry, an identical formula exists for the gauge field.
To see this, note that in Weyl gravity \cite{Ghilen1,Ghilen2}, see also eq.(\ref{non2}) in the text,
non-metricity is different from  EP quadratic gravity:
 $\tilde\nabla_\lambda g_{\mu\nu}=-\w_\lambda\, g_{\mu\nu}$. Contracting
this equation with $g^{\mu\nu}$ and using  $\tilde\nabla_\lambda \sqrt{g}=(1/2) \sqrt{g} \,g^{\mu\nu} \,
\tilde\nabla_\lambda g_{\mu\nu}$ we find
$\w_\lambda=(-1/2) \tilde\nabla_\lambda \ln\sqrt{g}$.
 This is similar to Palatini case, eq.(\ref{w}) in 
the text, although the connection is different.
From  these last two equations, by writing the action of $\tilde\nabla_\lambda$ on $g_{\mu\nu}$ 
one immediately finds
\bea\w_\mu= (1/2) (\tGamma_\mu-\Gamma_\mu(g))\label{ppp}
\eea
with the trace of
Levi-Civita connection  $\Gamma_\mu(g)=\partial_\mu \ln\sqrt{g}$.
Eq.(\ref{ppp}) was used as  a definition for
 the gauge field in  Einstein-Palatini quadratic gravity, eq.(\ref{fmunu}).


\begin{thebibliography}{100}

{\small


\bibitem{E1}
A. Einstein,  ``Einheitliche Feldtheories von Gravitation und Electrizitat'', 
Sitzungber Preuss Akad. Wiss (1925) 414-419.

\bibitem{E2}
M. Ferraris, M. Francaviglia and C. Reina, ``Variational formulation 
of general relativity from 1915 to 1925,   ``Palatini's method'' discovered by Einstein in 1925'',
Gen. Rel. Grav. 14 (1982) 243-254.

\bibitem{c7} For a review and references, see 
  G.~J.~Olmo,
  ``Palatini Approach to Modified Gravity: f(R) Theories and Beyond,''
  Int.\ J.\ Mod.\ Phys.\ D {\bf 20} (2011) 413
  [arXiv:1101.3864 [gr-qc]].

\bibitem{So1} Another review  is:
  T.~P.~Sotiriou and S.~Liberati,
``Metric-affine f(R) theories of gravity,''
  Annals Phys.\  {\bf 322} (2007) 935
  [gr-qc/0604006].
  T.~P.~Sotiriou and V.~Faraoni,
  ``f(R) Theories Of Gravity,''
  Rev.\ Mod.\ Phys.\  {\bf 82} (2010) 451
  [arXiv:0805.1726 [gr-qc]].




\bibitem{cc1}
  H.~A.~Buchdahl,
  ``Representation Of The Einstein-proca Field By An A(4)*,''
  J.\ Phys.\ A {\bf 12} (1979) 1235.


\bibitem{c1}
  H.~A.~Buchdahl,
  ``Non-linear Lagrangians and cosmological theory,''
  Mon.\ Not.\ Roy.\ Astron.\ Soc.\  {\bf 150} (1970) 1.

\bibitem{Vollick2}
  D.~N.~Vollick,
  ``Einstein-Maxwell and Einstein-Proca theory from a modified gravitational action,''
  gr-qc/0601016.


\bibitem{Vollick1}
  D.~N.~Vollick,
``Born-Infeld-Einstein theory with matter,''
  Phys.\ Rev.\ D {\bf 72} (2005) 084026
  [gr-qc/0506091].

\bibitem{c2}
  S.~Cotsakis, J.~Miritzis and L.~Querella,
``Variational and conformal structure of nonlinear metric connection gravitational Lagrangians,''
  J.\ Math.\ Phys.\  {\bf 40} (1999) 3063
  [gr-qc/9712025].

\bibitem{c3}
  B.~Shahid-Saless,
  ``First Order Formalism Treatment of R + R**2 Gravity,''
  Phys.\ Rev.\ D {\bf 35} (1987) 467.


\bibitem{c4}
  E.~E.~Flanagan,
``Palatini form of 1/R gravity,''
  Phys.\ Rev.\ Lett.\  {\bf 92} (2004) 071101
  [astro-ph/0308111].

\bibitem{c5}
  E.~E.~Flanagan,
``Higher order gravity theories and scalar tensor theories,''
  Class.\ Quant.\ Grav.\  {\bf 21} (2003) 417
  [gr-qc/0309015].

\bibitem{c6}
  M.~Borunda, B.~Janssen and M.~Bastero-Gil,
``Palatini versus metric formulation in higher curvature gravity,''
  JCAP {\bf 0811} (2008) 008
  [arXiv:0804.4440 [hep-th]].

\bibitem{c8}
  L.~Järv, M.~Rünkla, M.~Saal and O.~Vilson,
``Nonmetricity formulation of general relativity and its scalar-tensor extension,''
  Phys.\ Rev.\ D {\bf 97} (2018) no.12,  124025
  [arXiv:1802.00492 [gr-qc]].

\bibitem{Percacci3}
R.~Percacci and E.~Sezgin,
``New class of ghost- and tachyon-free metric affine gravities,''
Phys. Rev. D \textbf{101} (2020) no.8, 084040
[arXiv:1912.01023 [hep-th]].
\bibitem{c9}
  L.~Querella,
``Variational principles and cosmological models in higher order gravity,''
  gr-qc/9902044.

\bibitem{VV0}
  V.~Vitagliano, T.~P.~Sotiriou and S.~Liberati,
 ``The dynamics of generalized Palatini Theories of Gravity,''
  Phys.\ Rev.\ D {\bf 82} (2010) 084007
  [arXiv:1007.3937 [gr-qc]].

\bibitem{SO1}
G.~Allemandi, A.~Borowiec, M.~Francaviglia and S.~D.~Odintsov,
``Dark energy dominance and cosmic acceleration in first order formalism,''
Phys. Rev. D \textbf{72} (2005), 063505
[arXiv:gr-qc/0504057 [gr-qc]].

\bibitem{Kozak}
  A.~Kozak and A.~Borowiec,
 ``Palatini frames in scalar–tensor theories of gravity,''
  Eur.\ Phys.\ J.\ C {\bf 79} (2019) no.4,  335
  [arXiv:1808.05598 [hep-th]].

\bibitem{Annala}
  J.~Annala,
  ``Higgs inflation and higher-order gravity in Palatini formulation'',
  PhD thesis, Univ of Helsinki,
 https://inspirehep.net/literature/1799417
  

\bibitem{Weyl1}
Hermann Weyl, Gravitation und elektrizit\"at, Sitzungsberichte der
K\"oniglich Preussischen Akademie der Wissenschaften zu Berlin (1918), pp.465;
Einstein's critical comment appended (on atomic spectral lines changes).

\bibitem{Weyl2}
Hermann Weyl ``Eine neue Erweiterung der Relativitätstheorie'' 
(``A new extension of the theory of relativity''), Ann. Phys. (Leipzig) (4) 59 (1919), 101-133.

\bibitem{Weyl3}
Hermann Weyl ``Raum, Zeit, Materie'', vierte erweiterte Auflage. Julius Springer, Berlin 1921
``Space-time-matter'', translated from German by Henry L. Brose, 1922, Methuen \& Co Ltd, London.


\bibitem{Scholz}
For  review and references on Weyl's theory, see  E.~Scholz,
``The unexpected resurgence of Weyl geometry in late 20-th century physics,''
  Einstein Stud.\  {\bf 14} (2018) 261
  [arXiv:1703.03187 [math.HO]];

\bibitem{Ghilen2}
  D.~M.~Ghilencea,
  ``Spontaneous breaking of Weyl quadratic gravity to Einstein action and Higgs potential,''
  JHEP {\bf 1903} (2019) 049
  [arXiv:1812.08613 [hep-th]].

\bibitem{Ghilen1}
  D.~M.~Ghilencea,
``Stueckelberg breaking of Weyl conformal geometry and applications to gravity,''
  Phys.\ Rev.\ D {\bf 101} (2020) no.4,  045010
  [arXiv:1904.06596 [hep-th]].


\bibitem{Winflation}
  D.~M.~Ghilencea,
``Weyl R$^{2}$ inflation with an emergent Planck scale,''
  JHEP {\bf 1910} (2019) 209
  [arXiv:1906.11572 [gr-qc]].

\bibitem{Dirac}
  P.~A.~M.~Dirac,
``Long range forces and broken symmetries,''
  Proc.\ Roy.\ Soc.\ Lond.\ A {\bf 333} (1973) 403.
  doi:10.1098/rspa.1973.0070

\bibitem{Smolin}
  L.~Smolin,
 ``Towards a Theory of Space-Time Structure at Very Short Distances,''
  Nucl.\ Phys.\ B {\bf 160} (1979) 253.

\bibitem{Cheng}
  H.~Cheng,
  ``The Possible Existence of Weyl's Vector Meson,''
  Phys.\ Rev.\ Lett.\  {\bf 61} (1988) 2182.

\bibitem{Quiros}
  I.~Quiros,
  ``Scale invariant theory of gravity and the standard model of particles,''
 E-print arXiv:1401.2643 [gr-qc].

\bibitem{Fulton}
  T.~Fulton, F.~Rohrlich and L.~Witten,
  ``Conformal invariance in physics,''
  Rev.\ Mod.\ Phys.\  {\bf 34} (1962) 442.


\bibitem{JW}
  J.~T.~Wheeler,
``Weyl geometry,''
  Gen.\ Rel.\ Grav.\  {\bf 50} (2018) no.7,  80
  [arXiv:1801.03178 [gr-qc]].

\bibitem{Moffat1}
  M.~de Cesare, J.~W.~Moffat and M.~Sakellariadou,
  ``Local conformal symmetry in non-Riemannian geometry and the origin of physical scales,''
  Eur.\ Phys.\ J.\ C {\bf 77} (2017) no.9,  605
  [arXiv:1612.08066 [hep-th]].

\bibitem{Oh}
  H.~C.~Ohanian,
``Weyl gauge-vector and complex dilaton scalar for conformal symmetry and its breaking,''
  Gen.\ Rel.\ Grav.\  {\bf 48} (2016) no.3,  25
  [arXiv:1502.00020 [gr-qc]].

\bibitem{ghilen}          
  D.~M.~Ghilencea and H.~M.~Lee,
  ``Weyl symmetry and its spontaneous breaking in Standard Model and inflation,''
  arXiv:1809.09174 [hep-th].



\bibitem{P}
  A.~Barnaveli, S.~Lucat and T.~Prokopec,
``Inflation as a spontaneous symmetry breaking of Weyl symmetry,''
  arXiv:1809.10586 [gr-qc].

\bibitem{Moffat2}
  J.~W.~Moffat,
``Scalar-tensor-vector gravity theory,''
  JCAP {\bf 0603} (2006) 004
  [gr-qc/0506021].

\bibitem{Heisenberg}
  L.~Heisenberg,
``Scalar-Vector-Tensor Gravity Theories,''
  arXiv:1801.01523 [gr-qc].

\bibitem{Jimenez}
  J.~Beltran Jimenez, L.~Heisenberg, T.~S.~Koivisto,
``Cosmology for quadratic gravity in generalized Weyl geometry,''
  JCAP {\bf 1604} (2016) no.04,  046
  [arXiv:1602.07287 [hep-th]].

\bibitem{Jimenez2}
  J.~Beltran Jimenez, T.~S.~Koivisto,
``Spacetimes with vector distortion: Inflation from generalised Weyl geometry,''
  Phys.\ Lett.\ B {\bf 756} (2016) 400
  [arXiv:1509.02476 [gr-qc]].

\bibitem{Hill}
  C.~T.~Hill,
``Inertial Symmetry Breaking,''
  arXiv:1803.06994 [hep-th].

\bibitem{Scholz2}
  E.~Scholz,
``Higgs and gravitational scalar fields together induce Weyl gauge,''
  Gen.\ Rel.\ Grav.\  {\bf 47} (2015) no.2,  7
  [arXiv:1407.6811 [gr-qc]].

\bibitem{Tann}
  W.~Drechsler and H.~Tann,
 ``Broken Weyl invariance and the origin of mass,''
  Found.\ Phys.\  {\bf 29} (1999) 1023
  [gr-qc/9802044].

\bibitem{Dengiz}
  S.~Dengiz,
``A Note on Noncompact and Nonmetricit Quadratic Curvature Gravity Theories,''
  Turk.\ J.\ Phys.\  {\bf 42} (2018) no.1,  70
  [arXiv:1404.2714 [hep-th]].


\bibitem{Ross}
  P.~G.~Ferreira, C.~T.~Hill, J.~Noller and G.~G.~Ross,
``Scale-independent $R^2$ inflation,''
  Phys.\ Rev.\ D {\bf 100} (2019) no.12,  123516
  [arXiv:1906.03415 [gr-qc]].


\bibitem{tH0}
  G.~'t Hooft,
 ``Local conformal symmetry: The missing symmetry component for space and time,''
  Int.\ J.\ Mod.\ Phys.\ D {\bf 24} (2015) no.12,  1543001.


\bibitem{Turok}
  I.~Bars, P.~Steinhardt, N.~Turok,
 ``Local Conformal Symmetry in Physics and Cosmology,''
  Phys.\ Rev.\ D {\bf 89} (2014) no.4,  043515
  [arXiv:1307.1848 [hep-th]].

\bibitem{TH}
  G.~'t Hooft,
  ``Imagining the future, or how the Standard Model may survive the attacks,''
  Int.\ J.\ Mod.\ Phys.\  {\bf 31} (2016) no.16,  1630022.

\bibitem{tH1}
  G.~'t Hooft,
``Local conformal symmetry in black holes, standard model, and quantum gravity,''
  Int.\ J.\ Mod.\ Phys.\ D {\bf 26} (2016) no.03,  1730006.

\bibitem{sS1}
  M.~Shaposhnikov and D.~Zenhausern,
``Quantum scale invariance, cosmological constant and hierarchy problem,''
  Phys.\ Lett.\ B {\bf 671} (2009) 162
  [arXiv:0809.3406 [hep-th]].

\bibitem{Armillis}
  R.~Armillis, A.~Monin and M.~Shaposhnikov,
  ``Spontaneously Broken Conformal Symmetry: Dealing with the Trace Anomaly,''
  JHEP {\bf 1310} (2013) 030
  [arXiv:1302.5619 [hep-th]].

\bibitem{higgsdilaton}
  F.~Bezrukov, G~K.~Karananas, J.~Rubio and M.~Shaposhnikov,
  ``Higgs-Dilaton Cosmology: an effective field theory approach,''
  Physical Review D {\bf 87} (2013) no.9,  096001
  [arXiv:1212.4148 [hep-ph]].

\bibitem{Monin}
  F.~Gretsch and A.~Monin,
  ``Perturbative conformal symmetry and dilaton,''
  Phys.\ Rev.\ D {\bf 92} (2015) no.4,  045036
  [arXiv:1308.3863 [hep-th]].

\bibitem{D1}              
  D.~M.~Ghilencea,
   ``Quantum implications of a scale invariant regularization,''
  Phys.\ Rev.\ D {\bf 97} (2018) no.7,  075015
  [arXiv:1712.06024 [hep-th]];
  ``Manifestly scale-invariant regularization and quantum effective operators,''
  Phys.\ Rev.\ D {\bf 93} (2016) no.10,  105006
  [arXiv:1508.00595 [hep-ph]];
  ``One-loop potential with scale invariance and effective operators,''
  PoS CORFU {\bf 2015} (2016) 040
  [arXiv:1605.05632 [hep-ph]].

\bibitem{Lalak2}
  D.~M.~Ghilencea, Z.~Lalak and P.~Olszewski,
 ``Two-loop scale-invariant scalar potential and quantum effective operators,''
  Eur.\ Phys.\ J.\ C {\bf 76} (2016) no.12,  656
  [arXiv:1608.05336 [hep-th]].


\bibitem{D2}
  D.~M.~Ghilencea, Z.~Lalak and P.~Olszewski,
  ``Standard Model with spontaneously broken quantum scale invariance,''
  Phys.\ Rev.\ D {\bf 96} (2017) no.5,  055034
  [arXiv:1612.09120 [hep-ph]].

\bibitem{Kubo}
J.~Kubo, M.~Lindner, K.~Schmitz and M.~Yamada,
``Planck mass and inflation as consequences of dynamically broken scale invariance,''
Phys. Rev. D{\bf 100} (2019) no.1, 015037
[arXiv:1811.05950 [hep-ph]].

\bibitem{K1}
  R.~Foot, A.~Kobakhidze, K.~L.~McDonald and R.~R.~Volkas,
  ``Poincar\'e protection for a natural electroweak scale,''
  Phys.\ Rev.\ D {\bf 89} (2014) no.11,  115018
  [arXiv:1310.0223 [hep-ph]].

\bibitem{R1}             
  P.~G.~Ferreira, C.~T.~Hill and G.~G.~Ross,
  ``Scale-Independent Inflation and Hierarchy Generation,''
  Phys.\ Lett.\ B {\bf 763} (2016) 174
  [arXiv:1603.05983 [hep-th]].

\bibitem{FRH1}
  P.~G.~Ferreira, C.~T.~Hill and G.~G.~Ross,
  ``Inertial Spontaneous Symmetry Breaking and Quantum Scale Invariance,''
  arXiv:1801.07676 [hep-th].

\bibitem{FRH2}
  P.~G.~Ferreira, C.~T.~Hill and G.~G.~Ross,
 ``Weyl Current, Scale-Invariant Inflation and Planck Scale Generation,''
  Phys.\ Rev.\ D {\bf 95} (2017) no.4,  043507
  [arXiv:1610.09243 [hep-th]].

\bibitem{CW-BL}  
  E.~J.~Chun, S.~Jung and H.~M.~Lee,
``Radiative generation of the Higgs potential,''
  Phys.\ Lett.\ B {\bf 725} (2013) 158
   Erratum: [Phys.\ Lett.\ B {\bf 730} (2014) 357]
  [arXiv:1304.5815 [hep-ph]].

\bibitem{lebedev}  
  O.~Lebedev and H.~M.~Lee,
  ``Higgs Portal Inflation,''
  Eur.\ Phys.\ J.\ C {\bf 71} (2011) 1821
  [arXiv:1105.2284 [hep-ph]].

\bibitem{Lalak}
  Z.~Lalak and P.~Olszewski,
``Vanishing trace anomaly in flat spacetime,''
  Phys.\ Rev.\ D {\bf 98} (2018) no.8,  085001
  [arXiv:1807.09296 [hep-th]].

\bibitem{SO2}
E.~Elizalde, S.~D.~Odintsov and A.~Romeo,
``Manifestations of quantum gravity in scalar QED phenomena,''
Phys. Rev. D \textbf{51} (1995), 4250-4253
[arXiv:hep-th/9410028 [hep-th]].

\bibitem{SO3}
I.~Buchbinder, S.~Odintsov and I.~Shapiro,
``Effective action in quantum gravity,''
Published in Bristol, UK: IOP (1992) 413 p.


\bibitem{Strumia}
  A.~Salvio and A.~Strumia,
``Agravity,''
  JHEP {\bf 1406} (2014) 080
  [arXiv:1403.4226 [hep-ph]].
  A.~Salvio and A.~Strumia,
``Agravity up to infinite energy,''
  Eur.\ Phys.\ J.\ C {\bf 78} (2018) no.2,  124
  [arXiv:1705.03896 [hep-th]].

\bibitem{JJ2}             
D.~Iosifidis, A.~C.~Petkou and C.~G.~Tsagas,
``Torsion/non-metricity duality in f(R) gravity,''
Gen. Rel. Grav. \textbf{51} (2019) no.5, 66
[arXiv:1810.06602 [gr-qc]].


\bibitem{S1}
  E.~C.~G.~Stueckelberg,
``Interaction forces in electrodynamics and in the field theory of nuclear forces,''
  Helv.\ Phys.\ Acta {\bf 11} (1938) 299.


\bibitem{P1}
  R.~Percacci,
``Gravity from a Particle Physicists' perspective,''
  PoS ISFTG {\bf } (2009) 011
  [arXiv:0910.5167 [hep-th]].

\bibitem{P2}
  R.~Percacci,
``The Higgs phenomenon in quantum gravity,''
  Nucl.\ Phys.\ B {\bf 353} (1991) 271
  [arXiv:0712.3545 [hep-th]].

\bibitem{Latorre}
  A.~D.~I.~Latorre, G.~J.~Olmo and M.~Ronco,
``Observable traces of non-metricity: new constraints on metric-affine gravity,''
  Phys.\ Lett.\ B {\bf 780} (2018) 294
  [arXiv:1709.04249 [hep-th]].
\bibitem{Lobo}
  I.~P.~Lobo and C.~Romero,
``Experimental constraints on the second clock effect,''
  Phys.\ Lett.\ B {\bf 783} (2018) 306
  [arXiv:1807.07188 [gr-qc]].


\bibitem{CMB1}
  K.~N.~Abazajian {\it et al.} [CMB-S4 Collaboration],
``CMB-S4 Science Book, First Edition,''
  arXiv:1610.02743 [astro-ph.CO].
https://cmb-s4.org/

\bibitem{CMB2}
  J.~Errard, S.~M.~Feeney, H.~V.~Peiris and A.~H.~Jaffe,
``Robust forecasts on fundamental physics from the foreground-obscured, 
gravitationally-lensed CMB polarization,''
  JCAP {\bf 1603} (2016) no.03,  052
  [arXiv:1509.06770 [astro-ph.CO]].


\bibitem{CMB3}
  A.~Suzuki {\it et al.},
``The LiteBIRD Satellite Mission - Sub-Kelvin Instrument,''
  J.\ Low.\ Temp.\ Phys.\  {\bf 193} (2018) no.5-6,  1048
  [arXiv:1801.06987 [astro-ph.IM]].

\bibitem{Englert} For an early study of this theory, see
  F.~Englert, E.~Gunzig, C.~Truffin and P.~Windey,
``Conformal Invariant General Relativity with Dynamical Symmetry Breakdown,''
  Phys.\ Lett.\  {\bf 57B} (1975) 73.

\bibitem{Hi}
P.~W.~Higgs,
``Quadratic lagrangians and general relativity,''
Nuovo Cim. \textbf{11} (1959) no.6, 816-820

\bibitem{book}
D. Gorbunov, V. Rubakov, 
``Introduction to the theory of the early Universe'', World Scientific, 2011.

\bibitem{Nakayama}
  A.~Edery and Y.~Nakayama,
  ``Palatini formulation of pure $R^2$ gravity yields Einstein gravity with no massless scalar,''
  Phys.\ Rev.\ D {\bf 99} (2019) no.12,  124018
  [arXiv:1902.07876 [hep-th]].

\bibitem{J1}
  R.~Jackiw and S.~Y.~Pi,
``Fake Conformal Symmetry in Conformal Cosmological Models,''
  Phys.\ Rev.\ D {\bf 91} (2015) no.6,  067501
  [arXiv:1407.8545 [gr-qc]].

\bibitem{J2}
  R.~Jackiw and S.~Y.~Pi,
``New Setting for Spontaneous Gauge Symmetry Breaking?,''
  Fundam.\ Theor.\ Phys.\  {\bf 183} (2016) 159
  [arXiv:1511.00994 [hep-th]].


\bibitem{tHooft}
  G.~'t Hooft,
  ``Local Conformal Symmetry: the Missing Symmetry Component for Space and Time,''
  arXiv:1410.6675 [gr-qc]. (Essay written for the Gravity Research Foundation - 2015 Awards for 
Essays on Gravitation).

\bibitem{K}
  C.~Kounnas, D.~Lüst and N.~Toumbas,
  ``R$^2$ inflation from scale invariant supergravity and anomaly free superstrings with fluxes,''
  Fortsch.\ Phys.\  {\bf 63} (2015) 12
  [arXiv:1409.7076 [hep-th]].

\bibitem{LAG}
  L.~Alvarez-Gaume, A.~Kehagias, C.~Kounnas, D.~Lüst and A.~Riotto,
``Aspects of Quadratic Gravity,''
  Fortsch.\ Phys.\  {\bf 64} (2016) no.2-3,  176
  [arXiv:1505.07657 [hep-th]].


\bibitem{VV}
  D.~N.~Vollick,
  ``Modified Palatini action that gives the Einstein-Maxwell theory,''
  Phys.\ Rev.\ D {\bf 93} (2016) no.4,  044061
  [arXiv:1612.05829 [gr-qc]]. 

\bibitem{J}
  K.~i.~Kobayashi and T.~Uematsu,
``Nonlinear Realization of Superconformal Symmetry,''
  Nucl.\ Phys.\ B {\bf 263} (1986) 309.
  doi:10.1016/0550-3213(86)90119-7

\bibitem{gh}
J.~Beltrán Jiménez and A.~Delhom,
``Ghosts in metric-affine higher order curvature gravity,''
Eur. Phys. J. C \textbf{79} (2019) no.8, 656
[arXiv:1901.08988 [gr-qc]].

\bibitem{toc}
D.~M.~Ghilencea,
``Gauging scale symmetry and inflation: Weyl versus Palatini gravity,''
[arXiv:2007.14733 [hep-th]].


\bibitem{II1}
  F.~Bauer and D.~A.~Demir,
``Higgs-Palatini Inflation and Unitarity,''
  Phys.\ Lett.\ B {\bf 698} (2011) 425
  [arXiv:1012.2900 [hep-ph]].

\bibitem{II2}
  F.~Bauer and D.~A.~Demir,
  ``Inflation with Non-Minimal Coupling: Metric versus Palatini Formulations,''
  Phys.\ Lett.\ B {\bf 665} (2008) 222
  [arXiv:0803.2664 [hep-ph]].

\bibitem{II3}
  T.~Koivisto and H.~Kurki-Suonio,
  ``Cosmological perturbations in the palatini formulation of modified gravity,''
  Class.\ Quant.\ Grav.\  {\bf 23} (2006) 2355
  [astro-ph/0509422].

\bibitem{II4}
  S.~Rasanen and P.~Wahlman,
``Higgs inflation with loop corrections in the Palatini formulation,''
  JCAP {\bf 1711} (2017) 047
  [arXiv:1709.07853 [astro-ph.CO]].

\bibitem{II5}
  V.~M.~Enckell, K.~Enqvist, S.~Rasanen and E.~Tomberg,
  ``Higgs inflation at the hilltop,''
  JCAP {\bf 1806} (2018) 005
  [arXiv:1802.09299 [astro-ph.CO]].

\bibitem{II6}
  T.~Markkanen, T.~Tenkanen, V.~Vaskonen and H.~Veermäe,
 ``Quantum corrections to quartic inflation with a non-minimal coupling: metric vs. Palatini,''
  JCAP {\bf 1803} (2018) 029
  [arXiv:1712.04874 [gr-qc]].

\bibitem{II7}
  L.~Järv, A.~Racioppi and T.~Tenkanen,
 ``Palatini side of inflationary attractors,''
  Phys.\ Rev.\ D {\bf 97} (2018) no.8,  083513
  [arXiv:1712.08471 [gr-qc]].

\bibitem{Palatini1}
  I.~Antoniadis, A.~Karam, A.~Lykkas and K.~Tamvakis,
  ``Palatini inflation in models with an $R^2$ term,''
  JCAP {\bf 1811} (2018) 028
  [arXiv:1810.10418 [gr-qc]].

\bibitem{Palatini2}
  V.~M.~Enckell, K.~Enqvist, S.~Rasanen and L.~P.~Wahlman,
  ``Inflation with $R^2$ term in the Palatini formalism,''
  JCAP {\bf 1902} (2019) 022
  [arXiv:1810.05536 [gr-qc]].

\bibitem{Palatini3}
I.~D.~Gialamas and A.~Lahanas,
``Reheating in $R^2$ Palatini inflationary models,''
Phys. Rev. D \textbf{101} (2020) no.8, 084007
[arXiv:1911.11513 [gr-qc]].


\bibitem{planck2018}
  Y.~Akrami {\it et al.} [Planck Collaboration],
  ``Planck 2018 results. X. Constraints on inflation,''
  arXiv:1807.06211 [astro-ph.CO].

\bibitem{Sta}
  A.~A.~Starobinsky
``A New Type of Isotropic Cosmological Models Without Singularity,''
  Phys.\ Lett.\ B {\bf 91} (1980) 99
   [Phys.\ Lett.\  {\bf 91B} (1980) 99]
   [Adv.\ Ser.\ Astrophys.\ Cosmol.\  {\bf 3} (1987) 130].


\bibitem{Patrignani:2016xqp}
  C.~Patrignani {\it et al.} [Particle Data Group],
  ``Review of Particle Physics,''
  Chin.\ Phys.\ C {\bf 40} (2016) no.10,  100001.


\bibitem{Edholm}
  J.~Edholm,
  ``UV completion of the Starobinsky model, tensor-to-scalar ratio, and constraints on nonlocality,''
  Phys.\ Rev.\ D {\bf 95} (2017) no.4,  044004
  [arXiv:1611.05062 [gr-qc]] and references therein.

\bibitem{BH}
  R.~Kallosh, A.~D.~Linde, D.~A.~Linde and L.~Susskind,
``Gravity and global symmetries,''
  Phys.\ Rev.\ D {\bf 52} (1995) 912
  [hep-th/9502069].
}
\end{thebibliography}
\end{document}